\documentstyle[12pt,epsf]{article}
\newcommand {\nc} {\newcommand}
\newcommand\noi{\noindent}
 \newcommand\beq{\begin{equation}}
 \newcommand\eeq{\end{equation}}

 \def\beqn{\begin{eqnarray}}
 \def\eeqn{\end{eqnarray}}

\nc{\re} {\mathop{\mathrm{Re}}} \nc{\im} {\mathop{\mathrm{Im}}} 
\nc{\case}[2] {\mbox{$\frac{#1}{#2}$}} 

  \nc {\gap} {\\[1ex]}
  
  \nc {\f} {\frac}                \nc {\s} {\sqrt}
  
  \nc {\kap} {\kappa}		
  \nc {\amp}[1] {\phi_{#1}}
  \nc {\sgmtot} {\sigma_{\rm tot}}
  \nc {\aml}[2] {\phi_{#1}^{\mathrm{#2}}}
  \nc {\text}[1] {\mbox{\small{#1}}}		
  \nc {\delsigL} {\Delta\sigma_{_{\mathrm L}}}  
  \nc {\delsigT} {\Delta\sigma_{_{\mathrm T}}}  
	 \def \pom {I\hspace{-1.4mm}P}
		\def \smpom {I\hspace{-1.1mm}P}

\begin{document}
\begin{titlepage}
\title{\begin{flushright}{\small BNL-HET 98/46\\ \vspace{-.15in} \small
CPT-98/P.3693 \vspace{.25in}} 
\\ \end{flushright}
\Large \bf The spin
dependence of high energy \\ proton scattering.}

\author{\small \bf{N.~H.\ Buttimore} \\{\small \it School of
Mathematics, University of Dublin, Trinity College,}\\ {\small \it Dublin 2,
Ireland}
\thanks{Supported in part by funds provided under the International
Collaboration Programmes IC/96/053 and IC/97/061 of Forbairt, Ireland}
\and {\small \bf B.~Z. Kopeliovich}\\ {\small \it Max-Planck-Institute f\"{u}r
Kernphysik, Postfach 103980,}\\ {\small \it 69029, Heidelberg, Germany}
\and {\small \bf E. Leader}\\ {\small \it Birkbeck College, University of
London,}\\ {\small \it Malet Street, London WC1E 7HX, England}
\and {\small \bf J.~Soffer}\\ {\small \it Centre de Physique
Th\'{e}orique-CNRS-Luminy,
Case 907 }\\{\small \it F-13288 Marseille Cedex 09-France}
\and {\small \bf T.~L. Trueman}\\ {\small \it Physics Department, Brookhaven
National Laboratory,}\\ {\small \it Upton, N.Y. 11973, U.S.A.}
\thanks{This manuscript has been authored under contract number
DE-AC02-98CH10886 with the U.S. Department of Energy. Accordingly, the U.S.
Government retains a non-exclusive, royalty-free license to publish or
reproduce the published form of this contributions, or allow others to do so,
for U.S. Government purposes.}}
\date{\small December 15, 1998}
\maketitle
\normalsize
\vspace{-.5in}
\begin{center}
{{\em This paper is dedicated to the memory of our}\\ {\em friend and
colleague Richard Slansky.}}
\end{center}
\newpage

\begin{abstract}
Motivated by the need for an absolute polarimeter to determine the beam
polarization for the forthcoming RHIC spin program, we study the spin
dependence of the proton-proton elastic scattering amplitudes
at high energy and small momentum transfer.
In particular, we examine experimental evidence for the
existence of an asymptotic part of
the helicity-flip amplitude $\phi_5$
which is not negligible relative
to the largely imaginary average non-flip amplitude
\(
         \phi_+  =  \case{1}{2}(\phi_1 + \phi_3)
\).
 We discuss theoretical estimates of \(
        r_5 = m\,\phi_5/\s{-t}\,\im \phi_+
\)
based upon several approaches: extrapolation
of low and medium energy Regge phenomenological results to high
energies, models based on a hybrid of perturbative QCD and non-relativistic
quark models, and models based on eikonalization techniques. We also apply
the rigorous, model-independent methods of analyticity and unitarity. We
find the preponderence of evidence at currently available energy indicates
that $r_5$ is small, probably less than $10\%$. The best available
experimental limit comes from Fermilab E704: combined with rather weak
theoretical assumptions those data indicate that $|r_5| < 15\%$. These
bounds are important because rigorous methods allow much larger values.
Furthermore, in contradiction to a widely-held prejudice that $r_5$
decreases with energy, general principles allow it to grow as fast as
$\ln{s}$ asymptotically, and some of the models we consider show an even
faster growth in the RHIC range. One needs a more precise measurement of
$r_5$ or to bound it to be smaller than $5\%$ in order to use the classical
Coulomb-nuclear interference technique for RHIC polarimetry. Our results show
how important the measurements of spin dependence at RHIC will be to our
understanding of proton structure and scattering dynamics.  As part of this
study, we demonstrate the surprising result that proton-proton elastic
scattering is self-analysing, in the sense that all the helicity amplitudes
can, in principle, be determined experimentally at small momentum transfer
without a knowledge of the magnitude of the beam and target polarization.
\end{abstract}
\end{titlepage}
\newpage

\section{Introduction} The need to understand the spin dependence 
of scattering
amplitudes at high energy and small momentum transfer is important 
for two   distinct
reasons. Firstly it is a great challenge to strong interaction theory, 
since it
involves the application of QCD in a kinematical region where 
non-perturbative
effects are important. QCD has had great success in the perturbative 
region, but
experiments at HERA at very small $x$ are already raising questions 
for which the
standard perturbative approach may be inadequate \cite{HERA}; 
and future experiments
at RHIC and LHC will produce a vast amount of data outside the 
perturbative region.
It is hard to imagine a global solution to a non-perturbative 
QCD effect such as
small-$t$ spin dependence, but it is becoming more and more 
urgent to try to make
some progress in this direction.  

Secondly the extremely important RHIC spin program \cite{RHIC, Guryn}, 
which will test many
elements of QCD at a new level of accuracy and detail, relies 
heavily upon an
accurate knowledge of the beam polarization. For the purpose of
measuring the beam
polarization $P$, the  Coulomb-Nuclear Interference 
(CNI) polarimeter is very
attractive: it has a reasonably large analyzing power 
(about 4\%) in a region of
momentum transfer ($|t| \approx 0.002 - 0.003 \mbox{ GeV}^2$) where the rate is
extremely high. This method depends on the dominance of the 
interference of the
one-photon exchange helicity-flip amplitude (by an abuse of the term, normally
called a Coulomb amplitude, more properly the magnetic amplitude) with the
non-flip strong hadronic amplitude, which is determined by the total cross
section. The accuracy of the method is limited by  our uncertain knowledge of
the hadronic helicity-flip amplitude; its interference with the {\em non-flip
one-photon exchange} amplitude has the same shape in this
$t$-region
\cite{schwinger,BGL,kz,trueman,bs} and so must be known, or limited in
size,  in order to achieve
the  required accuracy. The   requirements of RHIC polarimetry 
($\Delta P/P \leq
0.05$) \cite{polarimetry requirement} put very stringent demands on our
knowledge of the helicity-flip amplitude. This problem was the 
impetus that drew
our attention to the long-standing question of the size of the  
proton-proton
helicity-flip amplitudes.
	
This is not intended to be a paper on polarimetry, though 
we will inevitably make
further comments on the subject as appropriate;  indeed, 
the demands of RHIC just
cited set a standard for our investigation. The aim of the paper 
is to provide a
reliable assessment of what is known about the helicity-flip 
amplitudes and what is
expected for them at high energies on the basis of  
various approximate or rigorous
theoretical calculations.

Another well-known practical issue arising from our lack of knowledge of
spin dependence is in the determination of the total cross section via the use
of unitarity and the extrapololation of the differential cross section
\cite{ schwinger,Amaldi,BSW75}. In particular, this may lead to 
an overestimate of the
total cross section by an amount proportional to the ratio of 
the sum of the squares of the
helicity-flip amplitudes to the square of the non-flip amplitude at $t=0$. 
To put this statement
more correctly and more precisely, in well-known notation 
which will be fully
defined in Section 2, it will be overestimated by the factor \cite{bsw}
\begin{equation}
\sqrt{1 + \beta^2}
\end{equation} where
\begin{equation} \label{eq:zeta2}
 	\beta^2
 		=
		\frac{1}{4} \left( \frac{\Delta\sigma_{\rm L}} 
 	{\sgmtot} \right)^2
 	\f{\left( 1 + \rho_-^2 \right)}{\left( 1 + \rho^2 \right)}
 	+
 	\frac{1}{2} \left( \frac{\Delta\sigma_{\rm T}} 
 	{\sgmtot} \right)^2
 	\f{\left( 1 + \rho_2^2 \right)}{\left( 1 + \rho^2 \right)}.
\end{equation} 
Martin \cite{Martin/Marseille} has emphasized that, 
because this is a
ratio of squares, a quite good comparison between cross-sections obtained by
this  technique and more direct
measurements of $\sgmtot$ leaves room for substantial spin dependence.

Both of these experimental issues along with the 
theoretical studies using unitarity
and dispersion relations emphasize the importance of 
understanding spin dependence at
very small $|t|$. In addition, the very powerful tool 
of interference between Coulomb
and strong amplitudes for extracting small parameters 
(like the $\rho$ parameter for
unpolarized elastic scattering) is effective in this region.

Of course the interest of this physics has been understood 
for a very long time. The
earliest studies relevant to our work date from the sixties. 
Associated in large part
with the polarized proton programs at Argonne, CERN and Serphukov, 
there was a very
large amount of phenomenological work in the seventies, and there 
were at the same
time a number of new, fundamental ideas introduced. In the eighties 
and later, QCD
has led to new techniques for modeling the spin dependence 
of high energy
scattering, and the experimental program at Fermilab has 
made important
contributions in this field. Specific 
citations will be given at
the appropriate place in the following sections. With the coming 
of RHIC, the
experimental motivation is very strong to revisit past studies 
and to attempt
to make some advances on them. That is our purpose here.

Section 2 will lay the groundwork for subsequent discussion by 
defining the basic
amplitudes and expressing the various measurable polarization 
dependent quantities in
terms of them.  The general forms near
$t=0$ will be discussed using Regge concepts, especially charge 
conjugation $C$ and
signature
$(-1)^J$ of the exchanged system, and the implications for the 
asymptotic phase of
the various amplitudes. The terms ``pomeron" and ``froissaron" 
will be defined for
our purposes, and several general results will be reviewed. 

In Section 3 our best knowledge regarding helicity-flip amplitudes will 
be given. This
includes low and moderate energy Regge and  amplitude analysis for 
$pp$ and $\pi p$
scattering, the energy dependence of $P=A_N$ at small $t$ and the 
most pertinent piece of experimental information:  the measurement by E704 at
Fermilab of $A_N$ in the CNI region.

Section 4 applies the rigorous methods used to derive the 
Froissart-Martin  bound to
limit the energy dependence of the single  helicity-flip amplitude 
relevant for CNI,
and interprets this in terms of the impact parameter representation.

Section 5 contains a description and evaluation of several models which give
predictions for spin dependence at high energy. These will mainly address the
single helicity-flip amplitude relevant for the CNI polarimetry.

Section 6 reviews the issues of Coulomb enhancement and shows how, 
in principle, all the scattering amplitudes in
$pp$ scattering may be determined experimentally  {\em without 
knowledge of the beam
polarization P}. This method is contingent on being able to 
make measurements of very
likely tiny asymmetries and it may turn out not to be practical. Should such
determination prove to be practical,
 elastic $pp$ scattering could be used as a self-calibrating polarimeter.

 Lastly Section 7 gives our conclusions.

Before moving on to the body of the paper, we would like to say 
 that this work
originated at a workshop sponsored by the RIKEN BNL Research Center 
during the summer
of 1997 \cite{RIKEN}. During the workshop we, along with several other people, 
discussed and
analyzed various other methods of polarimetry. Some methods are very clean
theoretically and have good analyzing power; in particular, 
polarized hydrogen jet
targets provide a self-calibrating method
\cite{jet}, while elastic $ep$ scattering is calculable and has
 a very large analyzing power with longitudinal polarized electrons 
and transverse or
longitudinally polarized protons \cite{ep}. One can also calibrate 
an unpolarized
hydrogen target with a second low energy scattering off Carbon; this 
requires working
at larger
$|t|$ where the rate is much lower, but values of
$t$ for which the analyzing power is large are sure to exist, in 
particular in the
dip region 
\cite{BZK1}. Nuclear targets, either in colliding beam or fixed target 
modes, might
be useful for elastic scattering in the same way, using structure 
at larger $t$;
their use in the CNI region is subject to the same uncertainties as for $pp$ 
\cite{BZK1,bs2,KT}.  Finally, because the purely empirical asymmetry observed
in inclusive 
$\pi$ production is very large and the rate is high, it 
may be the most practical
initial polarimeter
\cite{inclusive}; it nearly meets the required precision standard 
but one needs data to calibrate this polarimeter using the same target and at the same
energy   (in the fixed target
mode) as will be used in RHIC. The choice of method obviously 
involves several
different kinds of factors some of which, such as technical and cost, are 
beyond the scope of this paper.

\section{Fundamentals and dynamical mechanisms} It has long been 
understood that the
measurement of helicity amplitudes at high energy could be a powerful 
tool for
determining the dynamical mechanisms for scattering in the asymptotic region
\cite{Gribov,MT1,10}; this is especially true for nucleon-nucleon
scattering because its very rich spin structure allows for a greater
variety of quantum numbers to be exchanged \cite{Volkov}.
Five independent helicity amplitudes are required to describe
proton-proton elastic scattering \cite{BGL,GGMW} :
\begin{eqnarray}
\phi_1(s,t) & = & \langle ++|M|++ \rangle , \nonumber \\ \nonumber
\phi_2(s,t) & = & \langle ++ |M|-- \rangle , \\ \nonumber
\phi_3(s,t) & = & \langle +- |M|+- \rangle , \\  \nonumber
\phi_4(s,t) & = & \langle +- |M|-+ \rangle , \\ 
\phi_5(s,t) & = & \langle ++ |M|+- \rangle .
\end{eqnarray}
Here we use the normalization of \cite{BGL}. Since we are interested 
only in very high energy $\sqrt{s}$, such as will be
available at RHIC, and very small
momentum transfer
$|t| < 0.05 \mbox{ GeV}^2$, we will generally neglect $m$ with respect to
$s$ and neglect $t$ with respect to $m$ to simplify the presentation of
the formulas which follow.  For example, $k^2 =
\sqrt{s(s-4m^2)}/4$ will be replaced by $s/4$.  Then
\begin{equation} \label{eq:sigmatot}
\sgmtot = \frac{4\pi}{s}\im (\phi_1(s,t) + \phi_3(s,t))|_{t=0} 
\end{equation}
and
\begin{equation}
\frac{d\sigma}{dt} = \frac{2\pi}{s^2} \{|\phi_1|^2 + |\phi_2|^2 + |\phi_3|^2 +
|\phi_4|^2 + 4|\phi_5|^2\}.
\end{equation}
We will also have occasion to discuss ({\it i}) scattering of unlike-fermions,
requiring a sixth amplitude
$\phi_6$, a single helicity-flip amplitude which degenerates to $-\phi_5$ 
for identical
particles (of course, $\bar{p} p$ elastic scattering requires only 5
amplitudes),  and ({\it ii})
scattering of a proton on a spin-zero particle, like a pion or a spinless
nucleus, requiring only two amplitudes, a non-flip and a flip amplitude.

We will consider only initial state polarization measurements. 
There are certainly
interesting things that can be said about final state 
polarizations, but the first
generation spin program at RHIC will not measure these and 
so we will not discuss
them here. Using only initial state polarization, with one or 
both beams polarized,
one can measure seven spin dependent asymmetries. We follow the 
notation of
\cite{BGL}. There are slight variations in the definitions used 
in the literature,
having to do with the orientation of axes.
\begin{eqnarray} \label{eq:asymdef}
A_N \frac{d\sigma}{dt}& =& -\frac{4\pi}{s^2} 
\im \{\phi_5^*(\phi_1 +
\phi_2 +
\phi_3 -\phi_4)\}, \nonumber \\ A_{NN} \frac{d\sigma}{dt}& =& 
\frac{4\pi}{s^2}
\{2|\phi_5|^2 + \re (\phi_1^*
\phi_2 -
\phi_3^* \phi_4) \}, \nonumber \\ A_{SS}  \frac{d\sigma}{dt}& =& 
\frac{4 \pi}{s^2}
\re \{\phi_1 \phi_2^* + \phi_3
\phi_4^*\}, \nonumber \\ A_{SL} \frac{d\sigma}{dt}& =& \frac{4 \pi}{s^2}   
\re \{\phi_5^* (\phi_1 +
\phi_2  -
\phi_3+ \phi_4)\}, \nonumber \\ A_{LL}  \frac{d\sigma}{dt}& =& 
\frac{2 \pi}{s^2}
\{|\phi_1|^2 +|\phi_2|^2 -|\phi_3|^2 - |\phi_4|^2\}.  
\end{eqnarray}
It will be convenient to introduce some shorthand:
\beq		\label{eq.phi-plus}
 	 \phi_+  =  \case{1}{2}(\phi_1 + \phi_3) \, ,
\qquad
 	 \phi_-  =  \case{1}{2}(\phi_1 - \phi_3) \, ,
\eeq
and
\beq \label{eq:rhos}
\rho_2 = \f{\re {\phi_2}}{\im {\phi_2}}, \qquad \rho_- = 
\f{\re {\phi_-}}{\im {\phi_-}}.
\eeq
There are also two cross section differences corresponding to 
longitudinal and
transverse polarization:
\beq
	\f{\im \phi_-(s,0)}{\im \phi_+(s,0)}
	=					\phantom{-}
	\f{1}{2} \f{\delsigL(s)}{\sgmtot(s)} \, ,
\quad
	\delsigL = \sigma_{^\to_\gets} - \sigma_{^\to_\to},
 					\label{longitud}	
\eeq
\beq
	\f{\im \phi_2(s,0)}{\im \phi_+(s,0)}
	=
	- \phantom{\f{1}{2}} \f{\delsigT(s)}{\sgmtot(s)} \, ,
\quad
	\delsigT = \sigma_{_{\uparrow\downarrow}}
	-
	\sigma_{_{\uparrow \uparrow}}
 					\label{tranvers}.	
\eeq
When the proton scatters elastically off a distinct spin
$1/2$ particle, there are two more measurable asymmetries: 
$A_{N}'$ and
$A_{LS}$, in obvious notation; these degenerate into $A_N$ and $A_{SL}$ 
respectively
when the two particles are identical. For scattering off a spin 
zero particle, there
is only one asymmetry which corresponds to $A_N$.

At these small values of $t$, the interference of the strong amplitudes
with the single photon exchange amplitudes will be important; this interference 
is central
to this paper. To lowest order in $\alpha$, the fine structure constant,  one
replaces
\beq \amp{i} \to 
        \amp{i} + \aml{i}{em}\exp(i\delta)
\eeq
	with hadronic and electromagnetic elements.
        The Coulomb phase $\delta $
is approximately independent of helicity \cite{BGL,cahn}
\beq
        \delta = \alpha\ln\f{2}{q^2(B + 8/\Lambda^2)} - \alpha\gamma
\eeq
        where $B$, often called ``the slope", is the logarithmic 
derivative of the
differential cross section at $t=0$, a number about $13\mbox{ GeV}^{-2}$
 and increasing
through the RHIC region, $q^2=-t$, 
       Euler's constant $\gamma =
        0.5772\ldots$ and $\Lambda^2 = 0.71\mbox{ GeV}^2$ reproduces
        the small momentum transfer dependence of the proton form
        factors assumed to satisfy
\beq
        G_E(q^2) = G_M(q^2)/\mu_p =(1 + q^2/\Lambda^2)^{-2}.
\eeq
For {\it pp} scattering at high $s$ and small $t$, the electromagnetic
	amplitudes are approximately
\beqn \label{eq:1 photon ex.}
	\aml{1}{em} &=& \aml{3}{em} = \f{\alpha s}{t}F_{1}{}^2 ,\nonumber
\gap
\aml{2}{em} &=& -\aml{4}{em} = \f{\alpha s \kappa^2}{4 m^2} F_2^2, \nonumber
\gap
	\aml{5}{em} &=& -\f{\alpha s \kappa}{2 m\s{-t}}F_1 F_2,
\eeqn
	where $\mu_p = \kap + 1$ is the proton's magnetic moment,
	and $m$ its mass. For the full expressions see, e.g., \cite{BGL}. 
The proton
electromagnetic form factors
	$F_1(q^2)$ and $F_2(q^2)$ are related to $G_E$ and $G_M$ \cite[section 12.2]{predazzi} by
\beq
	F_1  =  \f{G_E - G_M \, t/4m^2}{ 1 - t/4m^2} \, ,
\qquad
\kappa 	F_2  =  \frac{G_M - G_E}{ 1 - t/4m^2} \, .
\eeq
The relations between $\phi_1$ and $\phi_3$ and between $\phi_2$ and
$\phi_4$, Eq.\ (\ref{eq:1 photon ex.}), are special consequences of 
the quantum numbers of
the exchanged photon; they are not generally true for the full 
amplitudes. Relations of this type will be dealt with shortly.

Each hadronic amplitude $\phi_j$ can, in principle, be broken up 
into two parts
\begin{equation}
\phi_j\equiv \phi_j^R + \phi_j^{As}
\end{equation}
where $\phi_j^R$ is controlled by Regge pole 
type dynamics and, in our
normalization, decreases with energy roughly like $s^{-1/2}$ 
with respect to the 
asymptotic part $\phi_j^{As}$. Although the first term is essential to
understanding the data in the low-to-moderate energy region which overlaps
the RHIC range, we will focus here solely on the second term.

Consider first the dominant non-flip forward amplitude
$\phi_+$; this must have an asymptotic piece whose imaginary part 
grows with energy
as a consequence of its connection Eq.~(\ref{eq:sigmatot}) to 
the nucleon-nucleon total cross section.
There are two widely used forms for $\phi_+^{As}$ to describe the high energy
behavior of
$\sgmtot(pp)$,  which is flat
up to $\sqrt s\sim 20$ GeV, with a value of 38 mb and then 
grows to 43 mb at $\sqrt
s=63$ GeV increasing further to about 62 mb at the CERN
$Sp\bar pS$ collider ($\sqrt s=546$ GeV).
In the first, one fits the data with terms of the form  
$s \ln^p{s}$, $p \leq 2$ \cite{Gauron,Block}. This form is suggested by Regge
theory and the Froissart-Martin bound \cite{1}
\begin{equation} \label{eq:F.bnd}
\vert\phi_+\vert \leq c s \ln^2s \ \  \mbox{as} \ \  s\to\infty.
\end{equation} 
In this approach
$\im \phi_+^{As}$ receives contributions from the simple pomeron pole
$I\hspace{-1.4mm}P $, with intercept 
$\alpha_{\smpom}(0) = 1$, together with a contribution growing at the maximum
allowed rate
$s \ln^2s$ (sometimes referred to as a froissaron 
\cite{Gauron})
\begin{equation} \label{eq:F.grth} \im \phi_+^{As}(s)= 
a_{\smpom}\ s +a_{F}\ s \ln^2s.
\end{equation}
In the second, one introduces an ``effective'' pole, the
Landshoff-Donnachie pomeron \cite{2},
with $\alpha_{\smpom} = 1+\Delta_{\smpom}$, where typically $\Delta_{\smpom}
\sim 0.08$.   The ensuing behavior
\begin{equation} \label{eq:L-D.grth} \im \phi_+^{As}\propto 
s^{1 +\Delta_{\smpom}} \label{eq:dl}
\end{equation} 
gives an excellent description of the behavior of $\sgmtot(pp)$
and $\sgmtot(\bar{p} p)$ and many other reactions.  
This form is also suggested by perturbative QCD calculations \cite{3},
but with a larger value of $\Delta_{\smpom}$. However, ultimately, it
violates Eq.\ (\ref{eq:F.bnd}) and so must be modified at higher values of
$s$. This sort of behavior was obtained much earlier in QED-like theories
\cite{chengwu} where consistency with Eq.\ (\ref{eq:F.bnd}) was achieved
through eikonalizing the form Eq.\ (\ref{eq:dl}). The unitarization by
multi-pomeron exchange of a ``bare" pomeron which grows as
$s^{1+\Delta_{\smpom}}$,
$\Delta_{\smpom}>0$, is obtained by eikonal methods in
\cite{Volkovitsky,Dubovikov}; in those papers the relation of this result,
via unitarity, to multiplicity distributions and inclusive inelastic cross
sections is demonstrated. The resulting behavior is consistent with the
Froissart-Martin bound, Eq.~(\ref{eq:F.bnd}) but the approach to the
limiting asymptotic form is much more complex than is assumed in Eq.\
(\ref{eq:F.grth}). See the discussion later in Sections 4 and 5 and
references cited there regarding the eikonalization method.

There is also theoretical evidence, from a study of three-gluon 
exchange in QCD
\cite{5}, for a crossing-odd contribution to $\phi_+^{As}$ which 
grows with
energy slightly less rapidly than the pomeron exchange, and which would lead
to a very slow decrease of the quantity
\newline $(\sgmtot(pp)-\sgmtot(\bar{p}p))/(\sgmtot(pp)+\sgmtot(\bar{p}p))$ at
asymptotic energies. However, phenomenological studies of this so-called
odderon $O$ contribution \cite{Block,6}  suggest that in the RHIC energy
range its contribution is very small compared to the crossing-even part of
$\phi_+^{As}$. Roughly
\begin{equation}
\frac{\vert\phi_+^{As}\vert^{odd} }{\vert\phi_+^{As}\vert^{even} }\le 2\%
\end{equation}
in the RHIC region and we shall therefore neglect the crossing-odd
contribution to $\phi_+$ in what follows.

The key question for us is, do any of the non-dominant amplitudes 
$\phi_2$,
$\phi_-$ and, especially, $\phi_5$ have asymptotic behavior
characteristic of the pomeron or froissaron? There is abundant
evidence at low energy, some of which we will discuss in Section 3,
that these amplitudes fall off with energy with respect to
$\phi_+$ as one would expect from lower lying Regge-exchange. It is not known,
however, whether asymptotically  they have a small but non-zero ratio
to $\phi_+$. To characterize these amplitudes we will define relative 
amplitudes in
the following way:
\beqn \label{r def} r_2 &=R_2 + \imath I_2=& \f{\phi_2}{2\im {\phi_+}}, 
\nonumber \\ r_-
&=R_- + \imath I_-=& \f{\phi_-}{\im {\phi_+}}, \nonumber 
\\ r_5 &=R_5 + \imath I_5=&
\f{m\ \phi_5}{\sqrt{-t}\ \im{\phi_+}}, \nonumber \\ 
r_4 &=R_4 + \imath I_4=& - \f{m^2
\phi_4}{t\ \im {\phi_+}}.
\eeqn
Notice the factor 2 in the definition of $r_2$ which is there to 
simplify  many later formulas. The factors involving $t$ which have been
extracted reflect the fact that as $t \to 0$ the strong amplitudes 
$\phi_1$, $\phi_2$ and $\phi_3$  go to a possibly non-zero constant while
$\phi_4
\propto t$ and
$\phi_5 \propto \sqrt{-t}$
as a consequence of angular momentum conservation.
The various $r$'s will be
assumed to be complex and to vary with energy but their variation 
with $t$ over
the small region we consider will usually be neglected. 
See, however, Section 6.

The determination of the asymptotic spin dependence can be used to 
help identify the
dynamical mechanisms at work at high energy. We can classify the 
dynamical mechanisms
according to the the quantum numbers parity ($P$), charge conjugation ($C$) and
signature 
$(\tau)$ of the
$t$-channel exchange.  An amplitude $A_\tau$ is called even or odd 
under crossing
according as $\tau=+1$ or $-1$, since
\begin{equation} A_\tau (e^{i\pi}s,t)=\tau A_\tau^* (s,t).
\end{equation}

For nucleon-nucleon scattering there are three classes of exchanges
\cite{10,11} and they contribute to the amplitudes as shown in Table 1.

\begin{table}[h]
\centering
$\begin{array}{|c|c|c|}
\hline
\mbox{Class 1} & \mbox{Class 2} & \mbox{Class 3} \\ 
\tau = P = C	& \tau = -P = -C  &
\tau = -P = C
\rule[-.5cm]{0cm}{10mm} \\ \hline \phi_+,\phi_5, \phi_2 \!-\! \phi_4 &
\phi_- &
\phi_2 + \phi_4 \rule[-.5cm]{0cm}{10mm}\\ \hline I\hspace{-1.4mm}P, O,
\rho, \omega,  f, a_2 & a_1 &
\pi, \eta, b
\rule[-.5cm]{0cm}{10mm}\\ \hline 
\end{array}$
\caption{\sl Classification of $pp$ amplitudes by exchange symmetries 
and the
associated Regge poles}
\end{table}
\noi If the asymptotically dominant contribution has definite quantum numbers,
then unitarity requires that it has the quantum numbers of the vacuum
\cite{PT} ;  this is the
defining property of the pomeron. Note that it is the quantum number $C$ 
which
determines the relative sign of the contribution of a given exchange to
nucleon-antinucleon scattering i.e.
\begin{equation} A^{\bar pp}_{\tau,P,C}(s,t) = C\, A^{pp}_{\tau P,C} (s,t).
\end{equation}
This implies that pomeron dominance and the absence of an odderon requires not only that the
total  cross sections
for $p p$ and $\bar{p} p$ be equal, but also their real parts, or
$\rho$ values. Because the pomeron has $\tau$ = +1, the well-known 
argument relating
the phase of a scattering amplitude to its energy dependence, see e.g.
\cite{Eden},  tells us
that, if the asymptotic behavior of $(\sigma_{pp}+ \sigma_{\bar{p} p})$
goes as
$ s^{\alpha -1}
\ln^p{s}$, then the amplitude for $C=+1$ exchange  goes as 
$ s^\alpha
\ln^p{s}\,\exp(-\imath
\alpha \pi/2) (1 -
\imath \, p\, \pi/2\ln{s})$. Either of the two behaviors Eq.\
(\ref{eq:L-D.grth}) or Eq.\ (\ref{eq:F.grth}) imply that at the maximum RHIC
energy range
\begin{equation}
\rho^{As}\equiv \frac{\re \phi_+^{As}}{\im \phi_+^{As}}\approx 0.12,
\end{equation}
but the energy dependence over the entire range is somewhat different.
(Of course,
a detailed fit over the {\it entire} RHIC range will require the 
inclusion of lower
lying Regge trajectories.)

It is not known whether the pomeron couples to $\phi_5$ or to
$\phi_2-\phi_4$.  The phenomenological success at medium energies of 
``$s$-channel
helicity conservation'' \cite{s-channel} would suggest a small coupling, 
but this question is open to
experimental study. If they do couple to the pomeron they will 
have exactly the same asymptotic phase  as $\phi_+$. This may prove useful in
investigating whether or not the dominant behavior becomes pure
pomeron/froissaron as $s \to \infty$, or  if there can be substantial
odderon contribution to these subdominant amplitudes.  
An odderon with nearly the same
asymptotic behavior as the pomeron/froissaron will be approximately
$\pi /2$ out of phase with it. As we have noted its coupling to
$\phi_+$ is quite weak, but nothing at all is known about its 
coupling to
$\phi_2 - \phi_4$ or $\phi_5$ and these phase relations may prove 
useful in probing
for such couplings. This matter is of great interest and is 
discussed in a separate
paper \cite{7}.

The exchanged objects with the quantum numbers assignments in
Table 1 could be pure Regge poles or cuts generated by the exchange
of the Regge pole plus any number of pomerons. These cuts will have
an asymptotic behavior which differs only by a power of $\ln{s}$ from
the simple Regge pole and so must be considered along with it
\cite{Mandelstam}. In general, although the couplings of pure poles
factorize, there is no reason for the cut couplings to do so. It is obvious
that the charge conjugation parity of a cut is equal to the product of that
of the poles that produce it. The corresponding situation with signature and
parity is less obvious because of the relative orbital angular momentum the
exchanged poles can have \cite{Gribov}. It has been shown, however,
\cite{Gribov1,Polkinghorne,Branson} that the signature of the cut
$\tau_{cut} =
\tau_{pole}$. This
means that the important relation between $C$ and $\tau$, that
distinguishes Classes 1 and 3 from Class 2 
in Table 1, is preserved for
the cuts. The situation for parity is not as certain; Jones and Landshoff
\cite{Jones} have shown that the ``wrong'' parity cut, $P_{cut}= -
P_{pole}$ is {\em suppressed} compared to the
``right'' parity cut,
$P_{cut}= + P_{pole}$. The strength of the suppression 
remains a quantitative question which is open 
to experimental and theoretical study. 

There are some very general things one can say about how the spin 
dependence can
help distinguish pole from cut contributions; for an early example 
see \cite{Gribov}.
If factorization should hold to a good approximation then one has
\beq
\phi_2(s,t)= - \f{\phi_5^2(s,t)}{\phi_+(s,t)} \quad \mbox{and} 
\quad \phi_-(s,t) = 0.
\eeq
This obviously leads to a very simple spin dependence. In particular it
implies, that as $t \to 0$, $\phi_2 \propto t$ 
rather than
the generally allowed behavior.

Even if factorization is not valid, some of the same conclusions 
can be obtained
just on the basis of quantum numbers.
One particularly important example has to do with
$\phi_2$ and $\phi_4$. We have had little to say about $\phi_4$ 
because angular
momentum conservation forces it to vanish linearly as $t \to 0$. If either
factorization holds or the dominant exchange has pure $CP=1$ or 
$CP=-1$, then
$\phi_2$ must also vanish in the forward direction \cite{MT1,PT}. The
first condition we have just seen. The second can be confirmed by
examining the table. There one sees that $\phi_2 +
\phi_4$ and $\phi_2 -\phi_4$ couple to opposite values of $CP$. Therefore if
only one value  of $CP$ is dominant
asymptotically, $\phi_2 \sim \mp \phi_4 \mbox{
as } s \to \infty$ and it, too, must vanish at $t=0$. This makes the
measurement of $\phi_2$ near $t=0$ a very interesting probe of the
dynamics; it may, at the same time have the unfortunate side effect
of making some asymmetries unmeasurably small.

Finally, notice from the table that neither the pomeron nor the
odderon have the quantum numbers required to couple to $\phi_-$;  
it thus seems unavoidable that
\begin{equation}
\Delta \sigma_{\rm L} = \f{16\pi}{s} \im \phi_-
\end{equation}
should vanish like $s^{-1/2}$ as $s\to\infty$. This we have seen is also a
consequence of factorization \cite{MT1}. If it does not, it indicates an
asymptotically important exchange other than the pomeron or the 
odderon. Such an
object has never been suggested to our knowledge, but there is 
no obvious reason that it should not exist.

We see here some very simple statements that one can make which
characterize the dynamics of high energy scattering by means of 
the spin variables.
If the dynamics is well approximated by a pure pomeron  pole 
the spin asymmetries
will be quite small and require very sensitive experiments to measure. 
One should
note that various suppressions as in pomeron vs. odderon or pole vs. cut
\cite{Jones} become gradually stronger (logarithmically with $s$ or as a very
small power of $s$); it will therefore be important to make these measurements
over as wide an energy range as possible. RHIC presents a wonderful opportunity
to do this.

\section{\bf  Best experimental knowledge of ${\bf \phi_-}$, ${\bf \phi_2}$ and
${\bf \phi_5}$}

As we have seen above, all the various spin observables are 
expressed in terms of the
helicity-flip amplitudes. Clearly, to achieve a full amplitude analysis, 
one needs a
substantial number of measurements, in the same kinematic region
which is,
unfortunately, far from the present experimental situation. 
Nevertheless, it is
possible to extract from the available data some very useful 
information on the
helicity-flip amplitudes which we will now try to review and summarize.

Among the different spin observables we will consider, the 
transverse single-spin
asymmetry $A_N$ (or ``analyzing power'') has been extensively 
measured for $pp$
elastic scattering, so it will play a central role in the 
following discussion.\\

{\it 3.1 $A_N$ in the $CNI$ region}\\

The only experiment which has obtained relevant data in this 
kinematic region \newline where
$-t$ is around $3 \times 10^{-3} \mbox{ GeV}^2$, is E704 at Fermilab 
\cite{E704} at a lab
momentum \newline $p_L=200 \mbox{ GeV}/c$; the results are shown in Fig.1, 
along with two curves
which will be explained shortly.
\begin{figure}[h]
\centerline{\epsfbox{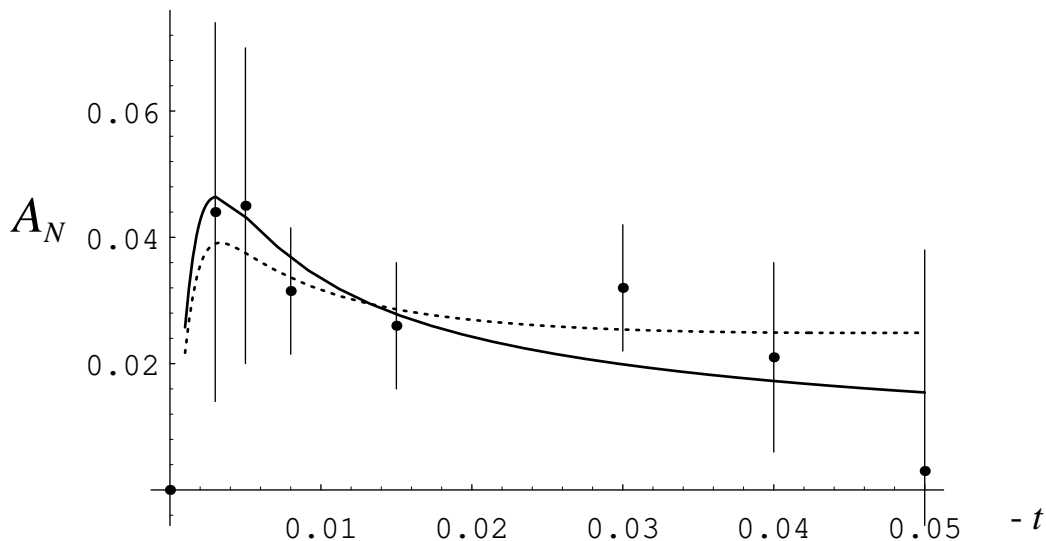}}
\medskip
\caption{\sl The data points are from Fermilab E704 \cite{E704}. The solid
curve  is the best fit
with the hadronic amplitude
$\phi_5$ constrained to be in phase with hadronic $\phi_+$ ; the dotted curve 
is the best fit
without this constraint.}
\end{figure}
 The errors are unfortunately too large to allow an unambiguous theoretical
interpretation, but let us now briefly recall what can one learn 
from it. From the
formulae in Section 2, $A_N$ is given by the expression 
(this is identical to the
expression for the final state polarization parameter $P$)

\begin{equation}\label{eq:A_N}
 	A_N
 	=
 { {	\im \lbrace(2\phi_+ + 2 e^{i\delta}\phi_+^{\rm em} + \phi_2)^\ast
\,
	(\phi_5 + e^{i\delta}\phi_5^{\rm em}) \rbrace
   }	\over
   {	|\phi_+ + e^{i\delta}\phi_+^{\rm em}|^2 + |\phi_-|^2
	+
	\frac{1}{2}|\phi_2|^2 + 2|\phi_5 + e^{i\delta}\phi_5^{\rm em}|^2
 } }	\, \,  , 
\end{equation}
 for not too large values of $-t$, such that the amplitude
$\phi_4 = \langle + - | \phi | - + \rangle $
 may be ignored because of the kinematical factor $(-t)$ 
occuring in this double helicity-flip amplitude.

In the one-photon exchange approximation $\phi_+^{\rm em}$ and 
$\phi_5^{\rm em}$ are real and have
well established expressions Eq.\ (\ref{eq:1 photon ex.}), so in order to
make  a theoretical prediction
using Eq.\ (\ref{eq:A_N}), one needs to know the hadronic amplitudes
$\phi_+,\phi_-,\phi_2$ and
$\phi_5$. The imaginary part of the largest one, $\phi_+$, is related at
$t=0$ to the total cross section $\sgmtot$ and the interference between
$\phi_5^{\rm em}$ and $\phi_+$ is most prominent when $t=t_c$, where
$t_c=-8\pi\alpha/\sgmtot$.

The explicit expression can be obtained by substituting
the expressions from Section 2 into Eq.\ (\ref{eq:A_N}):
\beqn
	\f{m\,A_N}{\s{-t}} \, \f{16\pi}{\sgmtot^2}
\,
	\f{d\sigma}{dt} \, e^{-Bt}	\nonumber &	= &	
\left[ \kap \left(1- \delta \rho + \im r_2 - \delta
\re r_2 \right)-2\,
(\im r_5 - \delta \re r_5) \right] \f{t_c}{t}	\gap	\nonumber &&	-
2 (1 + \im r_2 ) \re r_5		+
2 (\rho + \re r_2) \im r_5, \\
\frac{16\pi}{\sgmtot^2}
\,
	\f{d\sigma}{dt} \, e^{-Bt} &	= &	\left( \f{t_c}{t} \right)^2	
-
	2(\rho + \delta)\f{t_c}{t}	+
\,	(1 + \rho^2)(1 + \beta^2)
\, ,					\label{seperate}
\eeqn
where $\beta$ is defined in Eq.~(\ref{eq:zeta2}).
	The asymmetry for the CNI region can thus be expressed \cite{nhb-bnl} as a
 quotient of a linear expression in $t_c/t$ in the numerator and
 a quadratic expression for $t_c/t$ in the denominator, neglecting terms of order $t$.

The Coulomb phase $\delta$ is small, about 0.02 in the CNI region, 
smaller at larger
$|t|$. It has a slight effect on the position of the maximum in $A_N$:
\begin{equation}
	\frac{t_{\mathrm max}}{t_c} = \sqrt{3}
	+
	\frac{8}{\kappa}(\rho \im r_5 - \re r_5)
	-
	(\rho + \delta)\, ,
\end{equation}
in the approximation where small quantities are kept to first order, 
but it enters
the numerator multiplied by small amplitudes and so can be  
neglected for $p p$ scattering. The height of the peak is mainly sensitive to
the unknown quantities
$\im r_2$ and $\im r_5$, while the shape  depends mainly on 
$\re r_5$. For example, an
$\im r_5$ value of $\pm 0.1$ modifies the maximum of $A_N$ by about 11\%.

There are two fits to the E704 data allowing a non-zero $r_5$ 
shown in Fig.1 \cite{trueman}; the other $r_i$'s are set to zero. 
The solid curve is the best fit subject to the constraint that
$\phi_5$ is in phase with $\phi_+$. The arguments in Section 2 
show that if $\phi_+$ and
$\phi_5$ have the same asymptotic behavior they will have the same
phase; in that case the best fit is $|r_5| = 0.0 \pm 0.16$. Fitting
without that constraint yields the dotted curve, which corresponds
$|r_5|= 0.2 \pm 0.3$ with a relative phase angle to $\phi_+$ of $0.15
\pm 0.27$ radians. Note the large uncertainties on these values. This is  
essentially the same
as an earlier fit obtained in
\cite{abp}. As emphasized in \cite{trueman,bs}, we see 
that a large value of $\im r_5$
generates a very large uncertainty on $A_{\rm max}$, which 
can be of the order of 30\% or
more.\\ 
 
{\it 3.2 Energy dependence of the spin flip amplitudes from nucleon-nucleon
scattering}\\

In the small $t$ region we have some miscellaneous data on their 
magnitude and energy
dependence. First, the transverse-spin total cross sections difference
$\Delta \sigma_{\rm T}$ is related to $\im r_2$ for $t=0$, according to 
$\im r_2=-\Delta
\sigma_{\rm T}/2 \sgmtot$. From the limited ZGS data \cite{WdeB}, 
we find that
$\im r_2$ decreases strongly in magnitude from $-6\%$ at $p_L=2\mbox{ GeV}/c$
to $-0.4\%$ at 
$p_L=6\mbox{ GeV}/c$.
One can speculate whether for higher energy, it will remain negative 
and small or
change sign and increase in magnitude. The charge exchange reaction 
$np\to pn$, can
be also used to evaluate the modulus of $\phi_2$ which dominates 
the cross section
near the forward direction. The analysis of the data \cite {FI}, 
leads to the value
$|r_2|=3.5\% $ at $p_L=25\mbox{ GeV}/c$ and $|r_2|=0.6\% $ at
$p_L=270\mbox{ GeV}/c$, further evidence for a strong energy fall off of the
$I=1$ exchange amplitude. 

The longitudinal-spin total cross sections difference
$\Delta \sigma_{\rm L}$ is related to $\im r_-$ for $t=0$, according to 
$\im r_-=\Delta
\sigma_{\rm L}/2\sgmtot$. From the ZGS data \cite{auer}, we find that
$\im r_-$ decreases strongly in magnitude, from $-10\%$ at 
$p_L=2\mbox{ GeV}/c$ to $-0.6\%$ at
$p_L=12\mbox{ GeV}/c$. At higher energy, E704 has measured $\Delta \sigma_{\rm
L}$
\cite{Grosnick} for both $pp$ and $\bar{p}p$; their values 
imply that $\im r_-$ has decreased below
$10^{-3}$ for $pp$ and to about $10^{-3}$ (with a 100 \% error) for 
$\bar{p}p$. These
findings are consistent with the belief that $\im \phi_-$ vanishes as 
$s \to \infty$.  
\begin{figure}
\centerline{\epsfbox{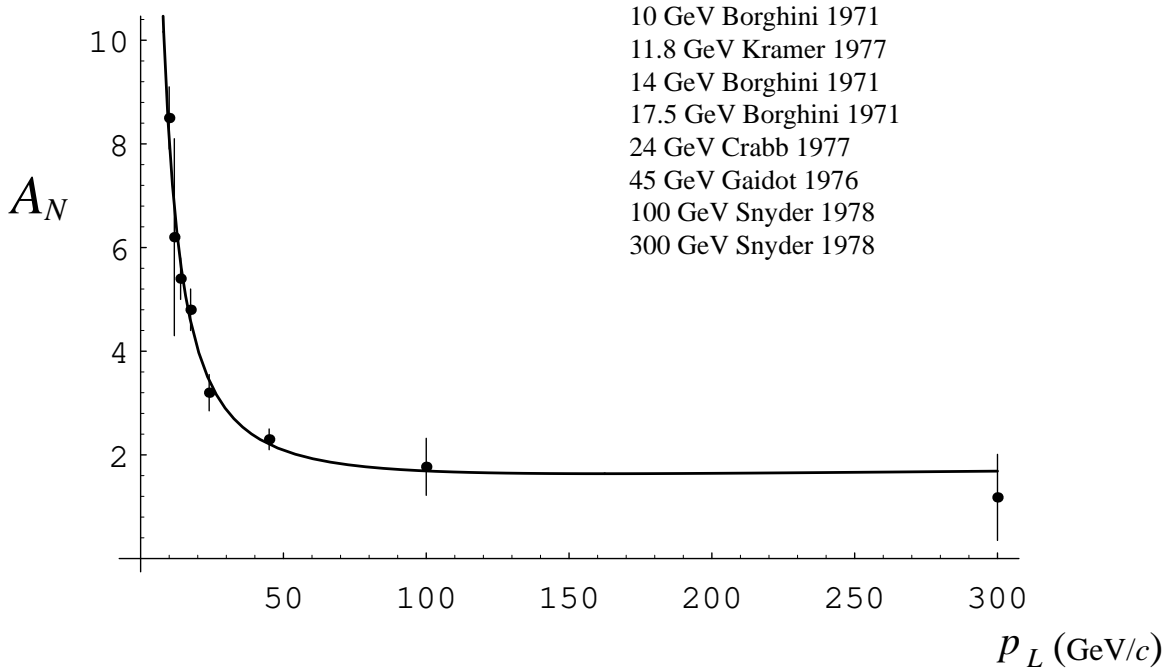}}
\medskip
\caption{\sl $A_N = P$ in percent as a function of lab momentum $p_L$ at
$t=-0.15\ GeV^2$}
\end{figure}

Away from the forward direction and the CNI region, the data 
indicate that
$A_N$ in $pp$ elastic scattering is falling very fast with energy. This 
has sometimes led to
the conclusion that the helicity-flip amplitude $\phi_5$ would 
vanish as a power of
$s$ as $s \to \infty$ \cite{krisch}. In order to investigate this, 
we
have taken a collection of data from various experiments which 
measure $P=A_N$ at different energies, all for
$t=-0.15\mbox{ GeV}^2$ (or interpolated from nearby values), the
smallest 
$|t|$ for which there is sufficient data to do this
\cite{data}. We have tried a fit suggested by Regge poles, namely
$P= a + b/\sqrt{p_L} + c/p_L$ \cite{trueman}. This is shown in Fig.~2 
and the relevant result is that
$a=0.023 \pm 0.012$.  It is not very well determined: it is 
consistent with pure CNI,
which is approximately 0.01 at this value of $t$ and $p_L \approx 300
\mbox{ GeV}/c$. At the same time it is consistent with a very large hadronic
helicity-flip amplitude: the calculated value of $A_N$ with
$\im r_5 = -0.6$ and $\re r_5 = -.015$ (so that $\phi_5$ is in phase with $\phi_1$)
approximates the fit very well for
$p_L$ above $200\ \mbox{GeV/c }$. Because of the phase energy relation discussed in
Section~2, these data are consistent with a large helicity-flip pomeron coupling. The
real and imaginary parts of $r_5$ cannot be separately determined from the measurement
of $A_N$ at this one value of $t$ , but they could both be determined by measuring the
$t$-dependence at RHIC because the deviation from the pure CNI {\it shape} is extremely
sensitive to $\re r_5$.\\

{\it 3.3 Iso-scalar part of the helicity-flip from $\pi^{\pm}p$ scattering}\\

Detailed Regge fits were made to spin dependent measurements in the 1970's
\cite{i&w, PDB, berger, bls}. At the low energies at which those
measurements were made,
there were quite large asymmetries observed. It was found that
these were mainly due
to the low-lying Regge trajectories and were not very sensitive
to the pomeron
couplings. The parameters that were found do predict a
very small ($< 10\%$) ratio of
the flip to non-flip residues for the pomeron, but the
parameters are uncertain
because of this insensitivity of the fits.

Polarization in
$\pi p$ elastic scattering at high energy is mostly due to
interference of the pomeron
non-flip amplitude with the helicity-flip part of the $\rho$-Reggeon.
As a consequence,
the polarization has different signs and is nearly symmetric in
$\pi^{\pm}p$
scattering. It	decreases with energy as
\beq A_N^{\pi p}(s,t)\propto
\left(\frac{s}{s_0}\right)^ {\alpha_{\rho}(t)-\alpha_{\smpom}(t)}\ ,
\label{1}
\eeq
\noi where $\alpha_{\rho}(t)\approx 0.5+0.9\ t$ and
$\alpha_{\smpom}(t)\approx 1.1+0.25\ t$. The polarization has a double-zero
behavior at
$t\approx -0.6 \mbox{ GeV}^2$, which is correlated to the change of sign of
$\alpha_{\rho}(t)$ at this point; see Fig.~3. This effect can be understood
as a  result of
destructive interference with the
$\rho\otimes \pom$ cut. An alternative explanation involves the wrong
signature  nonsense zero
\cite{PDB} (zeros in the residue and in the signature factor of the
$\rho$-reggeon).

At very high energies this part of  the polarization vanishes,
and one can hope to
detect an energy-independent contribution of the pomeron.
Unfortunately, available
data are not sufficiently precise yet.
One can eliminate the large  background from the $\rho\otimes \pom$
contribution by adding the
data on polarization in $\pi^{\pm}p$ elastic scattering,
\beq
\Sigma_{\pi p}(s,t) =
\delta_{+}(s,t)\
\ A_N^{\pi^+p}(s,t) +
\delta_{-}(s,t)\
\ A_N^{\pi^-p}(s,t)\ ,
\label{2}
\eeq
\noi where
\beq
\delta_{\pm}(s,t)=\frac{2\ \sigma^{\pi^{\pm}p}_{\rm el}(s,t)}
{\sigma^{\pi^{+}p}_{\rm el}(s,t)
+\sigma^{\pi^{-}p}_{\rm el}(s,t)}\ ,
\label{2a}
\eeq
\noi and
\beq
\sigma^{\pi^{\pm}p}_{\rm el}(s,t)\equiv
\frac{d\sigma^{\pi^{\pm}p}_{\rm el}(s,t)}{dt} \approx
\frac{\sigma^{\pi^{\pm}p}_{\rm tot}(s)^2}{16\pi}\
\exp\left(B_{\rm el}^{\pi^{\pm}p}t\right)\ .
\label{2b}
\eeq
\noi Therefore, Eq.\ (\ref{2a}) can be rewritten as,
\beqn
\delta_{+}(s,t)&=&\frac{2\ \gamma(s,t)} {1+\gamma(s,t)}\ ,\nonumber\\
\delta_{-}(s,t)&=&\frac{2} {1+\gamma(s,t)}\ ,
\label{2c}
\eeqn
\noi where
\beq
\gamma(s,t)=\left(\frac{\sigma^{\pi^+p}_{\rm tot}}
{\sigma^{\pi^-p}_{\rm tot}}\right)^2\
e^{-\Delta B\,t}.
\label{2d}
\eeq
\noi The difference between the elastic slopes
$\Delta B\equiv B^{\pi^-p}_{el}-B^{\pi^+p}_{el}$ is related
to the position of the
cross-over point $t_0 \approx -0.15 \mbox{ GeV}^2 $, which is
nearly energy independent
\cite{ira1} in this energy range since
\beq
\Delta B(s) = \frac{2}{|t_0|}
\ln\left(\frac {\sigma^{\pi^-p}_{\rm tot}(s)} {\sigma^{\pi^+p}_{\rm tot}(s)}
\right).
\label{2e}
\eeq

To find $\delta_{\pm}(s,t)$ we fit the data on
$\sigma^{\pi^{\pm}p}_{\rm tot}(s)$ \cite{tot} with the expression
\beq
\sigma^{\pi^{\pm}p}_{\rm tot}(s) =
\sigma_{\smpom}\ \left(\frac{s}{s_0}\right)^{\alpha_{\smpom}(0)-1} +
\sigma_f\ \left(\frac{s}{s_0}\right)^{\alpha_f(0)-1} \mp
\sigma_{\rho}\ \left(\frac{s}{s_0}\right)^{\alpha_{\rho}(0)-1}.
\label{2f}
\eeq
We fixed $\alpha_{\smpom}(0)=1.1$, $\alpha_f(0)=\alpha_{\rho}(0)= 0.5$,
$s_0=1\,\mbox{GeV}^2$
and found $\sigma_{\smpom}=12.4  \pm 0.03
\mbox{ mb} ,\ \sigma_f =40.8 \pm 0.26\mbox{ mb},\ \sigma_{\rho}=5.1 \pm
0.07\mbox{ mb}$.

Due to the cancellation of the isovector terms in Eq.\ (\ref{2})
$\Sigma_{\pi p}(s,t)$ is
dominated by the interference of the pomeron with the leading
isoscalar reggeons. In
$\pi p$ scattering this can only involve
$f$-reggeon interference.

Since the main part of the polarization cancels in the sum,
the data have
to have sufficiently high statistics in order to use Eq.\ (\ref{2}).
This is why we
could only use the low-energy data at  momenta $p_L= 6 - 14\mbox{ GeV}/c$,
depicted in Fig.~3.
\begin{figure}[thb]
\centerline{\epsfbox{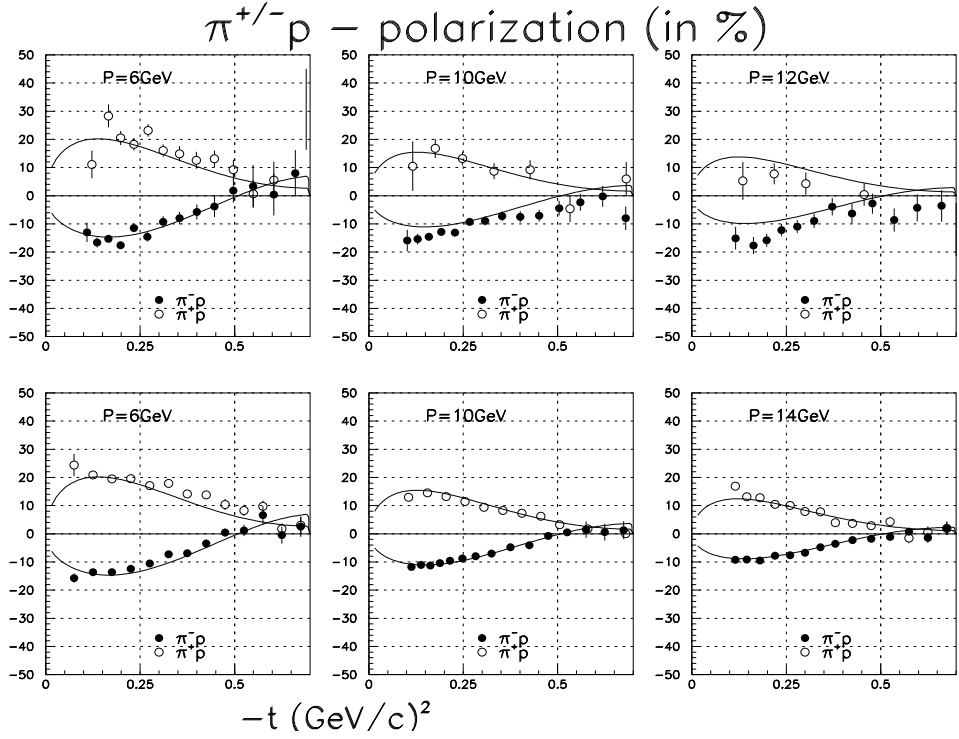}}
\medskip
\caption{\sl Polarization
in $\pi p$ elastic scattering in
the energy range $6 - 14\ GeV$. The data points are from
\cite{b67}-\cite{b71}. The curves show our fit with the parameterization
Eq.\ (\ref{3}) - Eq.\ (\ref{4}).}
\end{figure}

We performed a fit of $A_N^{\pi^{\pm}p}(s,t)$ with the parameterization
\beq A_N^{\pi^{\pm}}(s,t)= \frac {\Sigma_{\pi p}(s,t) \pm
\Delta_{\pi p}(s,t)} {2\ \delta_{\pm}(s,t)},
\label{2g}
\eeq
\noi where
\beq
\Delta_{\pi p}(s,t) = \frac{\sqrt{|t|}}{m_N}\
e^{a_1t}\,(a_2+a_3t)
(t-a_4)^2\ K(s)\,\left(\frac{s}{s_0}\right)^{\alpha_R(t)-\alpha_{\smpom}(t)}\ ,
\label{3}
\eeq
\noi and
\beq
\Sigma_{\pi p}(s,t) = \frac{\sqrt{|t|}}
{m_N}\ e^{a_1t}\,
(a_5+a_6t)\ K(s)\,\left(\frac{s}{s_0}\right)^{\alpha_R(t)-\alpha_{\smpom}(t)},
\label{4}
\eeq
\noi Here $t$ is in $(\mbox{GeV}/c)^2$.  We use the
same energy dependence for $\Sigma_{\pi p}(s,t)$ and
$\Delta_{\pi p}(s,t)$ assuming that
$\alpha_R(t)=\alpha_{\rho}(t) = \alpha_{f}(t)$.  The factor
\beq
K^{-1}(s)=
1+2\
\frac{\sigma_f}{\sigma_{\smpom}}\,\left(\frac{s}{s_0}\right)^{\alpha_R(t)
-\alpha_{\smpom}(t)} +\left\{\frac{\sigma_f}{{\rm
sin}\left[\frac{\pi}{2}\,\alpha_f(0)\right]
\sigma_{\smpom}}\,\left(\frac{s}{s_0}\right)^{\alpha_R(t)-
\alpha_{\smpom}(t)}\right\}^2
\label{5}
\eeq
takes into account the contribution of the $f$-reggeon to
the differential cross section.   The parameter $a_1$
corresponds to the difference between the slopes of the
pomeron and $f$-reggeon amplitudes. The factor $(t-a_6)^2$
is introduced to reproduce the double-zero behavior of the
polarization clearly seen in data \cite{b67}-\cite{b71}.
It is usually related
to presence of an additional zero in the $\rho$-reggeon
residue which is dictated by duality
at $\alpha_{\rho}=0$ \cite{PDB}.

The result of the fit is shown by the solid curves in
Fig.~3, and the values of the parameters $a_i$ are collected
in Table~2.

\begin{table}[h]
\centering
\begin{tabular}{|c|c|c|c|c|c|}
\hline
$a_1$ & $a_2$ & $a_3$ & $a_4$  & $a_5$&$a_6$ \\
\hline$ 2.5$&$12.5 $&$ 68.5 $&$ 0.7 $&$ 0.77 $&$ 2.33 $\\
$\pm .2$&$\pm 1.0$&$\pm 8.2 $&$ 0.01 $&$\pm 0.13 $&$ 0.59 $\\
\hline
\end{tabular}
\caption{\sl Fitted values of the parameters $a_i$}
\end{table}

The ratio of the helicity-flip to non-flip isosinglet amplitudes
can be extracted from
$\Sigma_{\pi p}(s,t)$,
\beq
r_f(t) - r_{\smpom}(t)={1\over 4}\,
e^{a_1t}\,(a_5+a_6t)
\frac{\sigma_{\smpom}}{\sigma_f} \tan
\left(\frac{\pi\alpha_f(t)}{2}\right).
\label{6}
\eeq
\noi Here $r_{\smpom} \sqrt{|t|/m^2}$ and $r_{f} \sqrt{|t|/m^2}$ denote the
ratio of the helicity-flip to non-flip amplitude corresponding to $\pom$
and $f$ exchange, respectively. We assume here that the helicity-flip and
non-flip amplitudes have the same phase which corresponds to dominance of
Regge poles.  We neglect the real part of the pomeron amplitude. If we
assume factorization, then for $pp$ scattering, asymptotically, $\im r_5(s,t)
\approx r_{\smpom}(t)$.

Thus, the combination Eq.\ (\ref{6}) of spin-flip to non-flip ratios for
iso-singlet amplitudes, which is quite difficult to measure
directly, are fixed by this analysis with a good accuracy.
\beq
r_f(0) - r_{\smpom}(0) = 0.06 \pm 0.01\ .
\label{7}
\eeq
Unfortunately, without further information regarding $r_f$, this does not
restrict $r_{\smpom}$. The approximation of
$f$-dominance of the pomeron yields $r_{\smpom}=r_f$, which obviously
contradicts Eq.\ (\ref{7}). The pion exchange model in Section 5.2
predicts values for both $r_{\smpom}$ and $r_f$ which are in pretty good
agreement with Eq.\ (\ref{7}). The Regge fits of \cite{PDB} use $r_f=0$.
This would give $r_{\smpom} = -0.06$, a very interesting value as we will
see in Section 5. However, this should probably be disregarded because the fits
of \cite{PDB}  also set the pomeron helicity-flip coupling to zero. 
The fits of
\cite{berger} give, assuming exchange degeneracy and using
the $\omega$ Regge residues, 
$r_f = 0.95/10.6 \approx 0.09$ and so, from Eq.\ (\ref{7}),
$r_{\smpom} = 0.03$. Since this result requires some theory that is not
tested to this precision, this can be taken as provisional but suggestive.

Another source of information on the isoscalar exchanges is $pA$
scattering. This requires special attention which we leave to another
occasion.

\section{Model-independent bounds and the energy dependence of helicity-flip}
The magnitude of $r_5$ depends on the scale $1/m$ chosen in 
Eq.\ (\ref{r def}),
where $m$ denotes the nucleon mass. This scale has been used 
conventionally for
many years; it was probably chosen in analogy to the form of the one-photon
exchange helicity-flip amplitude. It is not at all certain that this is the
appropriate scale for the scattering of strongly-interacting particles with
structure. It might be more natural for the scale to be set by the
slope of the diffraction peak; i.e. the effective radius of the proton
$R(s)=\sqrt{2 B(s)}$, (we take this to be the definition of the quantity
$R(s)$, see Eq.\ (\ref{eq:ftilde}) and Eq.\ (\ref{c}) below.) Since
this is a good deal larger than
$1/m$, the ``natural'' size of $r_5$ might be expected to be larger than 1.
Furthermore, it might very well be expected to increase slowly with
energy, corresponding to the growth in the effective radius of the
proton. This, of course, flies in the face of conventional wisdom; see the
discussion of Section 3.

It is natural to investigate if there is a theoretical argument that
$r_5 \rightarrow 0$ as $s \rightarrow \infty$. We begin by remarking
that for the pure Coulomb amplitudes this is not true, so we ask if
there is something different about the hadronic amplitudes. One obvious
difference is that experimentally $\phi_+$ grows faster than $s$,
and presumably will eventually grow as $s
\ln^2{s}$, the maximum rate allowed by the Froissart-Martin bound
\cite{1}. Let us see what the same arguments used to derive that bound yield
when applied to $\phi_5$. The partial wave expansion for $\phi_5$ 
is \cite{JW}
\begin{eqnarray}
\phi_5(s,t) &= &\sum_J (2J+1) f_J^5(s) d_{10}^J(\theta)  \nonumber \\
							&=& \f{\sin{\theta}}{2} \sum_J (2J+1) \sqrt{\f{J+1}{J}} f_J^5(s)
P_{J-1}^{(1,1)}(\cos{\theta}),
\end{eqnarray}
where $t = -2 k^2 (1-\cos{\theta})$ and $P_J^{(l,m)}(\cos{\theta})$ 
denotes the
Jacobi polynomial in $\cos{\theta}$ \cite{AG}. From this one finds that 
\begin{equation}
\hat{\phi_5}(s,0)= {m \over \sqrt{s}}\sum_J (2J+1) \sqrt{\f{J+1}{J}}f_J^5(s)
P_{J-1}^{(1,1)}(1),
\end{equation}
where $\hat{\phi_5} = m/\sqrt{-t} \ \phi_5$.
Partial wave unitarity requires that \cite{MS}
\begin{equation}
2|f_5^J(s)|^2 \leq \im {f_+^J(s)}
(1-\im{f_+^J(s)}) \leq 1/4 .
\end{equation}
 If we assume that this bound is
saturated out to some $L_{\rm max}(s) \sim k R(s)$, where $k \approx
\sqrt{s}/2$ is the  {\sc cm} momentum, then using
$P_{J-1}^{(1,1)}(1) = J$ (to be compared with $P_J(1)=1$ for the Legendre
polynomials) we find that for
$s \rightarrow \infty$,
$\hat{\phi_5}(s,0)$ goes as $m\, s\, R^3(s)$ while
$\phi_+$ goes as $s\, R^2(s)$, and so the natural scale for
$\hat{\phi_5}(s,0)$ {\em is} $R(s)$, not $1/m$. This means that unitarity and
other general principles allow $r_5$ to {\em grow} with energy; if the Froissart
bound is saturated $L_{\rm max} \sim \sqrt{s} \ln{s}$ and $r_5 \sim \ln{s}$ is
allowed. Note that if
$L_{\rm max} \sim \sqrt{s \ln{s}}$ then $\sgmtot$ will grow only as
$\ln{s}$ as favored by Block et al \cite{Block}; in that case, $r_5 \sim
\ln^{1/2}{s}$ is allowed. 

The above argument assumed the same $L_{\rm max}$ for $\phi_+$ and $\phi_5$.
This can, in fact, be proved as follows: one can bound
$P_J^{(1,1)}$ from below, parallel to Martin's argument for 
$P_l$, the Legendre
polynomial, in the unphysical region $|\cos{\theta}|> 1$. One then applies
the same reasoning as he used for $\phi_+$ to $\phi_5$. The representation
\begin{equation}
P_{J-1}^{(1,1)}(x) = \frac{2 J}{\pi} \int_{0}^{\pi} d \phi \ (x + \sqrt{x^2-1}
\cos{\phi}) ^{J-1} \sin^2{\phi} 
\end{equation}
allows one to show that $P_{J-1}^{(1,1)}(x) \sim  x^J/\sqrt{J}$ as
$J \rightarrow \infty $ for $x > 1$. This is the same as the asymptotic
behavior obtained for the $P_l(x)$ by Martin, and so polynomial
boundedness implies the same $L_{\rm max}$ for $\phi_5$ and $\phi_+$.

Notice that the same arguments applied to the double-flip amplitudes
$\phi_4(s,t)$ or $d\phi_2(s,t)/dt |_{t=0}$ will yield a natural
scale of
$R^2(s)$ and, correspondingly, a possible growth with energy as fast as
$\ln^4{s}$.

One can easily see that 
$\phi_5$ can
grow faster with $s$ than $\phi_+$ without violating unitarity because of the
factor of $\sqrt{-t}$. It is, naturally, an interesting question to
determine to what degree the  helicity-flip amplitudes saturate unitarity,
even at energies where the Froissart bound is not  saturated.
Techniques using unitarity and partial wave expansions have been used 
in the past
at low energy to obtain bounds on the helicity-flip amplitude in terms of
$\sgmtot$, $\sigma_{\rm el}$ and
$B$ \cite{mutter,hodgkinson,mennessier}; these bounds are comparable in
size to $mR/2$, i.e. $r_5$ is found to lie between 2 and 3.

We can make the discussion of unitarity more quantitative by transforming the
scattering amplitudes to the impact parameter representation. To keep the
discussion as simple as we can, let us do this for scattering of a proton on
a spin 0 target; as a $2 \times 2$ matrix, the scattering amplitude has the
form
\begin{equation} \label{eq:2by2}
\f{2 \pi}{\sqrt{s}}f(\vec{k'},\vec{k}) = g_1(s,q) + \vec{\sigma}  \cdot
\frac{\vec{k}
\times \vec{k'}}{|\vec{k}
\times \vec{k'}|} g_2(s,q),
\end{equation}
where $\vec{q} = \vec{k'}-\vec{k}$ , $q = |\vec{q}|$ and $q^2 = -t$ for
elastic scattering.

The two-dimensional Fourier transforms of these into impact parameter space
yields the profile functions $\tilde{g_1(b,s)}$ and $\tilde{g_2(b,s)}$:
\begin{equation}
\f{2 \pi}{\sqrt{s}}\int{\frac{d^2
\vec{q}}{2 \pi}\, e^{\imath
\, \vec{q} \cdot \vec{b}} f(\vec{k'},\vec{k})}
= \tilde{g_1}(b,s) + \imath \, \vec{\sigma}  \cdot
\frac{\vec{b} \times \vec{k}}{b k} \tilde{g_2}(b,s),
\end{equation}
where
\begin{eqnarray} \label{eq:f,g def}
\tilde{g_1}(b,s)  & = & \int{\frac{d^2 \vec{q}}{2 \pi}\,  e^{\imath \,
\vec{q} \cdot \vec{b}}} g_1(s,q), \nonumber \\
\tilde{g_2}(b,s)  & = & i \int{\frac{d^2 \vec{q}}{2 \pi}\, e^{\imath \,
\vec{q} \cdot \vec{b}} \,\hat{b} \cdot \hat{q}}\, g_2(s,q). 
\end{eqnarray}
With this normalization
\begin{equation} \label{eq:fnorm}
\sgmtot(s) = 4 \pi \int{b\ db \im {\tilde{g_1}(b,s)}},
\end{equation}
and unitarity imposes, for each value of $b$, the condition
\begin{equation} \label{eq:b_unitarity}
2 \im {\tilde{g_1}}(b,s) \geq |\tilde{g_1}(b,s)|^2 + |\tilde{g_2}(b,s)|^2. 
\end{equation}
(This equation is, in general, only approximate in $b$ space, but it can be
derived from the analogous partial wave inequality \cite{MS} if only the
{\em elastic}  scattering amplitudes are sufficiently peaked in $t$.) The
bounds discussed earlier correspond to a uniform distribution in
$b$ for both amplitudes for $b \leq R = L_{\rm max}/k$. If this
$b$-distribution is translated into the
$t$-dependence of the amplitudes near $t=0$ it implies that the slope of
$g_2/\sqrt{-t}$ is less than the slope $B=L_{\rm max}^2/2 k^2$ of $g_1$;
in fact it is
$3B/5$.

A more conventional assumption is that the
slopes of $g_1$ and $g_2/\sqrt{-t}$ are the same. If, in fact,
$g_2(s,q)=\lambda \ (q/m) \, g_1(s,q)$, with
$\lambda$ independent of $t$ then
\begin{equation}
\tilde{g_2}(b,s) =\frac{\lambda}{m} \frac{d\tilde{g_1}(b,s)}{db}.
\end{equation}
This is true in the optical model or in any other
model where the potential shape or matter distribution is the same for
spin-orbit force as for the purely central force. Then $\tilde{g}(s,b)$ will
be more peripheral than for the bound just discussed. It has
nothing intrinsically to do
with unitarity or  saturation of the Froissart bound, and it is clearly
interesting  to determine whether it is true or not. 

The unitarity condition Eq.\ (\ref{eq:b_unitarity})
imposes a bound on $|\lambda|$, and the closer $\tilde{g_1}(b,s)$ is to
saturating unitarity, the stronger this bound will be. Approximating the
$t$-dependence of the amplitude by a logarithmically shrinking diffraction peak
and neglecting its real part gives
\beq \label{eq:ftilde}
\tilde{g_1}(b,s)= \frac{i\,\sigma(s)}
{2\pi R^2(s)}\,{\rm exp}\left[-\frac{b^2}{R^2(s)}\right]\ ,
\eeq
and 
\beq \label{eq:gtilde}
\tilde{g_2}(b,s) = -\frac{2\ i b\, \lambda \,\sigma(s)}{2 \pi m R^4(s)}\,{\rm
exp}\left[-\frac{b^2}{R^2(s)}\right]\ ,
\eeq
where here and in the rest of this section the energy dependent Regge radius of
interaction is
\beq
R^2(s)=R_0^2 +4\alpha'_{\smpom} \,
{\rm ln}\left({s\over s_0}\right).
\label{c}
\eeq
With this form for the amplitudes $\sigma(s) = \sgmtot(s)$ via Eq.\
(\ref{eq:fnorm}). This will change at the next stage of the calculation. Here
one finds, numerically, over a wide range of values of $\sigma(s)/ 2 \pi
R^2(s) \leq 1$ that
\beq \label{eq:lambda bound}
|\lambda| \leq mR \sqrt{\f{2 \pi R^2(s)}{\sigma(s)}}
\eeq
is required in order to satisfy Eq.\ (\ref{eq:b_unitarity}).

If $\sigma(s)$ grows faster than $R^2(s)$ with $s$ as $s \to \infty$, say
as $s^{\Delta_{\smpom}}$ \cite{2}, then the amplitude Eq.\ (\ref{eq:ftilde})
will eventually violate the unitarity condition Eq.\ (\ref{eq:b_unitarity})
and the form must be modified.   It is well-known that the total $pp$ cross
section at available  energy is still far below the Froissart-Martin bound;
however, the bound Eq.\ (\ref{eq:b_unitarity}) is already saturated at small
impact parameters,  even ignoring the helicity-flip piece
\cite{Volkovitsky, as}. In principle,  unitarity is restored after all the
Regge cuts  generated by multi-pomeron exchanges are added
\cite{Dubovikov}. A standard way of unitarization of the non-flip part
of the pole amplitude \cite{PDB} is eikonalization; however, 
the presence of the
helicity-flip component may lead to problems with unitarity.
Indeed, an even number of repeating helicity-flip amplitudes contribute to
the non-flip part, but all of them grow as a power of 
energy and have the same sign. Therefore, eikonalization of 
the helicity-flip amplitude alone does not save unitarity, which can be restored
only after the absorptive corrections due to initial/final state
spin non-flip interactions are included. The resulting profile function reads,
\beq
\tilde{g_1}^{eik}(b,s) = 1-{\rm exp}[i\,\tilde{g_1}(b,s)] +
\left\{1 - {\rm cosh}\left[\frac{2\,i\,\lambda\,b}
{m\,R^2(s)}\,\tilde{g_1}(b,s)\right]\right\}\,
{\rm exp}[i\,\tilde{g_1}(b,s)].
\label{d}
\eeq
The first two terms on the {\it r.h.s.} of this equation
correspond to eikonalization of the non-flip part of 
Eq.\ (\ref{eq:2by2}).
They obey the unitarity bound at any $s$ and $b$. In the extreme asymptotic
region where $\sigma(s)$ in Eq.\ (\ref{eq:ftilde}) and Eq.\ (\ref{eq:gtilde})
is much greater than $R^2(s)$ then the
$b$-dependence has the form of a ``black disk'', {\it i.e.}
$\tilde{g_1}^{eik}(b,s) = 1$ at
$b < \widetilde R(s)$ and vanishes at $b>\widetilde R$, where 
\cite{Dubovikov}, 
\beq
\widetilde R^2 =
\Delta_{\smpom}\,{\rm ln}\left({s\over s_0}\right)\,R^2(s)\ ,
\label{e}
\eeq
if $\sigma(s) \sim (s/s_0)^{\Delta_{\smpom}}$. 
Likewise,
\begin{equation}
\tilde{g_2}^{eik} = - {\rm sinh}\left[\frac{2\,i\,\lambda\,b}
{m\,R^2(s)}\,\tilde{g_1}(b,s)\right] \,
{\rm exp}[i\,\tilde{g_1}(b,s)].
\end{equation}

Problems with unitarity at $b<\widetilde R(s)$ 
may arise from the last term in Eq.\ (\ref{d}). 
The condition Eq.\ (\ref{eq:b_unitarity}) is satisfied if
\beq
|\re \lambda| < \frac{m\,R^2(s)}{2\,b}
\label{f}.
\eeq
The minimal bound corresponds to a maximal $b=\widetilde R$,
and $s\to\infty$,
\beq
|\re \lambda | < m\,\left(\frac{\alpha'_{\smpom}}
{\Delta_{\smpom}}\right)^{1\over 2}.
\label{g}
\eeq
For reasonable values of $\alpha_{\smpom}'$ and $\Delta_{\smpom}$  we conclude
that $|\re \lambda| < 1.6$. This is not a severe restriction, 
and is valid only in the extreme asymptotic limit, beyond the RHIC range;
numerical calculations give a much larger bound, of order $mR$ at 
RHIC energies. 

Note that
$\lambda$ is renormalized by the eikonalization process; the result,
$\lambda^{eik}(s)$ can be calculated numerically from
\beq
\lambda^{eik}(s)= \frac{m \int{db\ b^2 \tilde{g_2}^{eik}(b,s)}}{2 \int{db\ b
\,
\tilde{g_1}^{eik}(b,s)}}.
\eeq
Likewise, the total cross section will be modified from the input values
$\sigma(s)$ and is given by
\begin{equation}
\sgmtot(s) =  4 \pi \int{b\ db \im \tilde{g_1}^{eik}(b)}.
\end{equation}
These last two equations will have to be used for comparison with data.

\section{Models for the pomeron helicity-flip}
An early attempt to understand the spin structure of the pomeron coupling was
made by Landshoff and Polkinghorne \cite{lands&polk}. This model preceded the
formulation of QCD, but used some of its features in a model they called the dual
quark-parton model. They argued that the $t$-dependence of the pomeron
coupling was determined by the electromagnetic form factors of the proton and
neutron. This led to the conclusion that the helicity-flip coupling is given
by the isoscalar anomalous magnetic moment of the nucleons; in our notation
$r_5=(\mu_p -1 +\mu_n)/2 = -0.06$. This relation has subsequently been
obtained or conjectured independently in a variety of models based on QCD.
The result is, however, model-dependent as we will see.

{\it 5.1 Perturbative QCD }

There is a widespread prejudice that the perturbative pomeron does not
flip helicity. It is  true that the perturbative pomeron 
couples to a hadron through
two $t$-channel gluons, and that the quark-gluon vertex $\bar
u_q\gamma_{\mu}u_q$ conserves  helicity. However, one cannot jump 
to the conclusion
that the same is true for a proton. In QED the fundamental vertex has
the same form but radiative corrections induce helicity-flip via an
anomalous magnetic moment. Ryskin \cite{ryskin} evaluated the pomeron
helicity-flip coupling by analogy to the {\em isoscalar} anomalous
magnetic moment of the nucleon. Applying this analogy to the quark gluon
vertex he found the anomalous color magnetic moment of the quark. Thus the
quark-gluon vertex does not conserve helicity and one can calculate the
helicity-flip part of the pomeron-proton vertex. Using the two-gluon model
for the pomeron and the nonrelativistic constituent quark model for the
nucleon he found
\cite{ryskin}
\begin{equation}
\im r_5 = 0.13,
\end{equation}
independent of energy. In the above one needs to introduce an effective gluon
mass and if one takes a large effective gluon mass,
$m_g
\approx 0.75 \mbox{ GeV}$, this estimate is substantially reduced. A need for
a large gluon mass follows from lattice QCD calculations \cite{Shuryak} and
the smallness of the triple-pomeron coupling \cite{triple-pom}.

The spin-flip part of the three-gluon odderon was also estimated in
\cite{ryskin} and the helicity-flip component was found to be nearly
the same as for the pomeron. If this is so then 
the odderon-pomeron interference contribution to $A_N$ vanishes. See Eq.\
(\ref{eq:asymdef}) and Table 3 in Section 6.

An alternative approach is to note that helicity is  defined relative to the
direction of the proton momentum, while the quark momenta  are oriented
differently. Therefore, the proton helicity may be different from  the  sum
of the quark helicities
\cite{kz}. The results of perturbative QCD calculations show 
that the helicity-flip
amplitude in elastic proton scattering very much correlates 
with the quark wave
function of the proton. Spin effects turn out to cancel 
out if the spatial
distribution  of the constituent quarks in the proton 
is symmetric \cite{kz,z}.
However, if a quark configuration containing a compact 
diquark ($ud$) dominates the
proton wave function, the pomeron helicity-flip part is nonzero. 
The more the proton wave
function is asymmetric, {\it i.e.} the smaller the diquark is, 
the larger is $\im r_5$ \cite{kz,z}. Its value in the CNI region of 
transverse momentum ranges from
$-0.05$ to $-0.1$ and even to $-0.15$ for the
diquark diameters 
$0.5$, $0.3$ and $0.2\,{\rm fm}$, respectively. The commonly accepted  
diquark size is $0.3
- 0.4\ {\rm fm}$; therefore, we conclude that
$|\im r_5|$ does not exceed $10\%$.

Note that there is a principal difference in sensitivity to 
the proton wave function
between the helicity-flip and the  non-flip components of the pomeron. 
The former probes
the shortest  interquark distances in the proton (diquark), but the latter is
sensitive  to the largest quark separation (due to color screening). 
Correspondingly, the virtuality of the gluons in the pomeron is 
higher in the
helicity-flip component since these gluons must resolve the diquark structure. 
This fact
may be considered as a justification for perturbative calculations 
for the helicity-flip
part, while their validity for the non-flip part is questionable.

High gluon virtuality in the helicity-flip pomeron leads to a 
steep energy dependence. A
prominent experimental observation at HERA is that the steepness 
of growth with
energy of the total virtual photoabsorption  cross section 
correlates with the photon
virtuality $Q^2$, {\it i.e.} with the $q\bar q$ separation in the hadronic
fluctuation of the  photon. Analyses of the data for the proton 
structure function
$F_2(x,Q^2)$ performed in \cite{kp} shows that for a quark separation 
of the order of
the mean diquark diameter one should expect the energy dependence
$\sim(s/s_0)^{0.2}$. This should be compared with the well known  
energy dependence
of the non-flip amplitude, $\sim (s/s_0)^{0.1}$. Therefore if the perturbative
QCD model is meaningful in this region we expect a negative
$\im r_5 $ with energy dependence $(s/s_0)^{0.1}$. This prediction can be
tested in future polarization experiments at RHIC whose energy ranges from
$s\approx 50 \mbox{ GeV}^2$ (with a fixed target) up to $25\times
10^4 \mbox{ GeV}^2$.
$\im r_5$ is expected to double its value in this interval.

Eventually this growth will cause the bound Eq. (\ref{g}) to be
violated. This occurs only at very high energy, well above the LHC energy, and
so it is not important for our considerations. Nevertheless, it would be
interesting to develop an eikonalization method that would lead to consistent
unitary amplitudes. We believe the eikonalization procedure developed
in \cite{knp} is the appropriate technique. When the elastic amplitude depends on
transverse separation between partons, as it does here, the measured amplitude
is the result of averaging over different transverse configurations:
\begin{equation}
\tilde{f}(b,s) = \langle \tilde{f}(b,s,\psi) \rangle_{\psi},
\end{equation}
where $\psi$ characterizes the transverse configuration and the averaging is
weighted by the probability to be in configuration $\psi$. Correspondingly,
eikonalization has to be done first for a given configuration $\psi$ and only
then averaged :
\begin{equation}
\tilde{f}^{eik}(b,s) = \langle \tilde{f}^{eik}(b,s,\psi) \rangle_{\psi}.
\end{equation}
For a given partonic configuration $\psi$ the energy dependence of the
helicity-flip and non-flip components must be the same since, as stated above,
it depends only on the transverse separations. Therefore, restriction Eq.\
(\ref{g}) applies except that the pomeron intercept depends on $\psi$ and
unitarity is satisfied for each $\psi$. However, the weight factors are
different for the helicity-flip and non-flip amplitudes and the averaging
results in a higher effective intercept for the helicity-flip component. The
detailed predictions of this procedure remain to be worked out.

{\it 5.2 Pion exchange model} 

\begin{figure}[tbh]
\centerline{\epsfbox{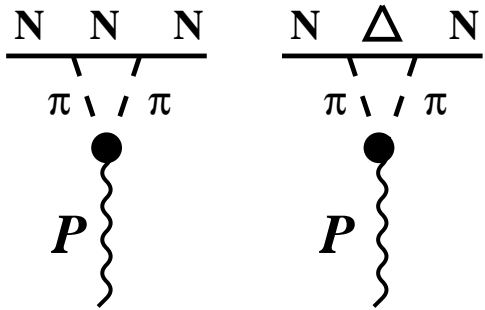}}
\medskip
\caption{\sl Pomeron coupling to a 
nucleon via two pion
exchange.}
\label{pion}
\end{figure}

A nucleon is known to have a pion cloud of large radius. 
Since the helicity-flip
amplitude is proportional to impact parameter, it is natural 
that a substantial
fraction comes from inelastic interaction of the projectile  
hadron with virtual
peripheral pions. This contribution is related through the 
unitarity relation to a
pomeron-nucleon vertex (in the elastic hadron-nucleon amplitude) 
shown in Fig.~\ref{pion}.
It is known that the main contribution to the pion cloud 
comes from the virtual
transitions $N\to \pi N$ and $N\to\pi\Delta$, which corresponds 
to the two graphs
depicted in Fig.~\ref{pion}. This model for the pomeron-nucleon 
coupling was suggested
in \cite{pumplin}. They predicted $\im r_5 \approx 0.016\,
(\ln{s})^{3/2}$.
This quite a steep energy dependence originates from the
radius of the  pion cloud
which is assumed to be proportional to $\sqrt{\ln{s}}$.  
A more 
detailed analyses was undertaken in \cite{pion}.  
An interesting observation of this
paper is a strong correlation of the value of $r_5$ 
with isospin in the $t$-channel.
Namely, for an isoscalar exchange ($\pom$, $f$-reggeon) the two graphs in
Fig.~\ref{pion} essentially cancel in the helicity-flip, but they add up in the
non-flip amplitude. It is vice versa for  an isovector exchange
($\rho$-reggeon).  This conclusion is consistent
with Regge phenomenological analyses of experimental data (see {\it e.g.}
\cite{regge}).

In order to fix the parameters of the model a detailed analysis 
of data on inclusive
nucleon ($pp\to p(n)X$) and $\Delta$ ($pp\to\Delta^{++}X$ and
$\pi^+p\to\Delta^{++}X$) production was performed in \cite{pion}. 
These reactions
correspond to the unitarity cut of the graphs in Fig.~\ref{pion}. 
The calculations in
\cite{pion} led to a positive value of $\im r_5 = 0.06$ 
for the pomeron (0.15 for
the $f$-reggeon). This nonperturbative contribution 
has the opposite sign to what
follows from perturbative calculations and may partially compensate  
it (see discussion in \cite{kz}).\\

{\it 5.3 Impact picture} 

An impact picture approach, which was derived 
several years ago \cite{BSW79,BSW84,BSW95},
describes successfully $\bar pp$ and $pp$ elastic 
scattering up to ISR energies . It led to predictions 
at very high energy, so far in excellent agreement 
with the data from the CERN SPS  collider and the
FNAL Tevatron and others, which remain to be checked at 
the Large Hadron Collider under construction at 
CERN . The spin-independent
amplitude reads at high energies
\begin{equation}
\phi_+^{impact}(s,t)=is\int^{\infty}_{0}J_0(b\sqrt{-t})(1-e^{-\Omega_0(s,b)})bdb~,
\end{equation}
where the opaqueness $\Omega_0$, which is assumed to factorize as 
$\Omega_0(s,b)=S_0(s)F(b^2)$, is associated
with the pomeron exchange. The energy dependence is given by the 
crossing symmetric function
\begin{equation}
S_0(s)= s^c/\ln^{c'}{s} + u^c/\ln^{c'}{u}~,
\end{equation}
which comes from the high energy behavior of quantum field theory. In
$S_0(s)$ above, $u$ is the third Mandelstam variable and both $s$ and $u$
are expressed in $\mbox{GeV}^2$. Note that $S_0(s)$ is complex because $u$ is
negative. The phenomenological analysis leads to the values of the
two free parameters $c = 0.167$, $c'= 0.748$ and the real part of
$\phi_+^{impact}(s,t)$ results from the phase of $S_0(s)$. The $t$-dependence
of
$\phi_+^{impact}(s,t)$ is driven by $F(b^2)$, which is related to  the Fourier
transform of the electromagnetic proton form factor and, as a result of a
simple parametrization which can be found in
\cite{BSW79},
$F(b^2)$ is fully determined in terms of only {\it four} additional
parameters.

\begin{figure}[htb]
\centerline{\epsfbox{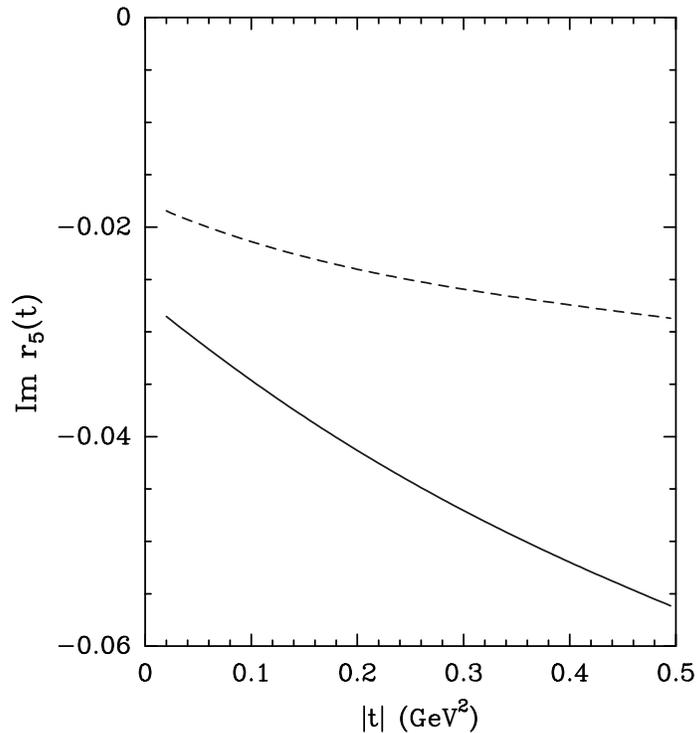}}
\medskip
\caption[flip]
{\sl $\im r_5(t)$ calculated in the impact picture for two energy values: 
$\sqrt s = 50\mbox{ GeV}$ (dashed curve) and $\sqrt s = 500 \mbox{ GeV}$
(solid curve) .}
\label{r5}
\end{figure}

The spin structure of the model was also studied and 
it allows a rather good description of the polarization data, 
up to the highest
available energy, {\it i.e.} $p_L=300 \mbox{ GeV}/c$ \cite{B82}. At
the RHIC energies, the spin dependent amplitude reads
\begin{equation}
\phi_5^{impact}(s,t)=is\int^{\infty}_{0}J_1(b\sqrt{-t})
\Omega_1(s,b)e^{-\Omega_0(s,b)}bdb~,
\end{equation}
where $\Omega_1(s,b)$ is the spin dependent opaqueness, 
corresponding to the helicity-flip component of the pomeron. It also 
factorizes as
$\Omega_1(s,b)=S_1(s)F_s(b^2)$, where $S_1(s)$ is obtained from 
$S_0(s)$ and we have
\begin{equation}
S_1(s)= {s^c \over \ln^{c'}{s}}(c - c'/\ln{s}) + (s \rightarrow u)~.
\end{equation}
$F_s(b^2)$ is simply related to $F(b^2)$ according to 
$F_s(b^2)=b\omega(b^2)F(b^2)$, where $\omega(b^2)$ is a smooth function which
is not very precisely known. The important point is its value $\omega_0$ for
very small $b$ and by fitting the data, it was found that 
$\omega_0=0.06\mbox{ GeV}$. This leads to a value $\im
r_5 \approx -0.06$, if one assumes that the
flip component of the pomeron is normalized at $t=0$, by the nucleon isoscalar
magnetic moment
\cite{BSW75}. This is at variance with the exact results one obtains in the
impact picture, which are shown in Fig.~\ref{r5} at two different energies.
It is interesting to remark that $\im r_5(t)$ increases with energy, in a way
pretty much consistent with what was mentioned above in Section 4.

\section{$P$-independent determination of ${\bf \phi_2}$,\ ${\bf
\phi_-}$,\ ${\bf
\phi_4}$ 
\mbox{and} ${\bf \phi_5}$}

In this section we would like to demonstrate that, in principle,
by making use of both CNI and hadronic interference at small
$t$ it is possible to determine all the spin dependent amplitudes
at $t=0$ {\em independent of knowledge of the beam polarization
$P_1$ and 
$P_2$} provided only that they are stable and non-zero. This is very
interesting, perhaps surprising, in its own right. If it proves to be
practical, it would permit the use of  elastic $pp$ scattering as a
self-calibrating polarimeter. It is important to emphasize right at the
beginning---we will not repeat  this every time the issue occurs---that the
method involves several ratios of very  small quantities; the precision required
to do this may be beyond the reach of practical experiment at this time.
However, very little is known about the amplitudes now, and so we
cannot evaluate this. Even if the complete process we describe
cannot be carried through, much of what follows should be useful
in constraining the amplitudes at small
$t$. 

The method requires the use of 
asymmetries with both longitudinal and transversely polarized beams; it will
not work unless data are available with both configurations. Here we work only
to order $\alpha$ and so only amplitudes that are large compared to the next
order correction can be determined from the formulas given below. This could
probably be improved upon if necessary; at the present, experiment will
probably not be able to probe amplitudes below that size and so we have not
pressed on in this direction. We assume that the polarized beams have the same
degree of polarization $P$ in either configuration; since they are produced
from the same initial configuration by rotation this is almost certainly true.
For simplicity in writing we assume both beams to have the same polarization;
this may very well not be true but it is trivial to correct the 
formulas for this.

We work with the experimentally measured asymmetries which are given by $P
A_N$, $P^2 A_{NN}$, etc. These will contain singular terms as $t \rightarrow
0$ coming from the interference between the one-photon exchange and the
hadronic amplitudes. To order $\alpha$ the asymmetries $A_{NN}, A_{SS}$ and
$A_{LL}$ are singular as $1/t$ and $A_N, A_{SL}$ are singular as $1/\sqrt{-t}$.
So we write

\begin{eqnarray}\frac{m\sqrt{-t}}{\sgmtot}\,P A_N \,
\frac{d\sigma}{dt} & = & - \alpha \, a_N +
\frac{\sgmtot}{8 \pi} \, b_N
\, t +\ldots ,  \nonumber \\
\frac{t}{\sgmtot}\,P^2 A_{LL} \, \frac{d\sigma}{dt} & = &
\alpha \, a_{LL} +
\frac{\sgmtot}{8 \pi} \, b_{LL}
\, t +\ldots ,\nonumber \\
\frac{t}{\sgmtot} \, P^2 A_{NN} \, \frac{d\sigma}{dt} & = &
\alpha \, a_{NN} +
\frac{\sgmtot}{8 \pi} \, b_{NN}
\, t +\ldots  ,\nonumber \\
\frac{t}{\sgmtot} \, P^2 A_{SS} \, \frac{d\sigma}{dt} & = &
\alpha \, a_{SS} +
\frac{\sgmtot}{8 \pi} \, b_{SS}
\, t +\ldots  ,\nonumber \\
 -\frac{m\sqrt{-t}}{\sgmtot}\, P^2 A_{SL} \,
\frac{d\sigma}{dt} & = & \alpha \, a_{SL} +    
\frac{\sgmtot}{8 \pi} \,b_{SL} \, t +\ldots .
\end{eqnarray}
In the following table we give expressions for the various
$a_i$, which we sometimes refer to as the enhanced  pieces, and
$b_i$ which we refer to as the hadronic piece. The
$a_i$'s are linear in the hadronic amplitudes while the
$b_i$ are bilinear. Here we omit terms of order $\alpha t$ which 
are small; we
will return to show how this can be corrected for, if necessary. 
Note that the usual exponential
$t$-dependence of the hadronic amplitudes will enter only at order $t^2$ or
$\alpha t$; likewise, for the quantities $\phi_5 ^2$ and $\phi_4$. In this
approximation $A_{NN} = A_{SS}$. For notation see Section 2.
\begin{table}[h]
\centering
$\begin{array}{|c|c|c|}
\hline
\mbox{asymmetry} & a_i & b_i \\ \hline A_N														&
P \{ \frac{\kappa}{2} (1 + I_2) - I_5\} &
P I_5 \{(\rho -
\rho_5) + I_2(\rho_2 - \rho_5)\}
\rule[-.5cm]{0cm}{10mm} \\ \hline A_{NN} & P^2 R_2 &
P^2 R_2 \{\rho +\frac{1}{\rho_2} + I_-(\rho_-
+\frac{1}{\rho_2})\} \rule[-.5cm]{0cm}{10mm}\\ \hline A_{LL} & P^2 R_-
& P^2 \{ R_-(\frac{1}{\rho_-} + \rho) + R_2 I_2
(\frac{1}{\rho_2} +\rho_2)\} \rule[-.5cm]{0cm}{10mm}\\ \hline
A_{SL} & P^2 \{\frac{\kappa}{2}(R_-+ R_2) \} & P^2 I_5 \{ (I_-
+
I_2) +\rho_5 (R_- + R_2) \}
\rule[-.5cm]{0cm}{10mm} \\ \hline
\end{array}$
\caption{\sl The first two terms in an expansion in $t$ of 
the various asymmetries.}
\end{table}

(The possibility of using the electromagnetic and hadronic pieces
of $A_{NN}$ and
$A_{LL}$ to determine the real and imaginary parts of $\phi_2$ 
and $\phi_-$,  when
the polarization is independently known was noticed in \cite{schwinger}.)

We also need the cross section differences

\begin{eqnarray} 
\Delta \sigma_{\rm T}& =& - 2 I_2 \, \sgmtot, \nonumber \\
\Delta \sigma_{\rm L}& =& 2 I_- \, \sgmtot.
\end{eqnarray}
Fits to the data will determine $ a_N,  b_N, a_{NN}, b_{NN}$, etc. The
strategy will be to take ratios of two quantities that are 
either linear or
bilinear in the polarizations to obtain ratios of amplitudes 
which will then be
independent of the polarization. We will find that there are
enough of these ratios to solve for all the amplitudes, provided
that at least one of
$\phi_2$ and $\phi_-$ is non-zero. Indeed, the system is over-constrained, 
and the procedure we describe here is not unique. We carry it 
through here to demonstrate that a solution exists;
 the optimal method will no doubt depend on
the experimental situation. If both $\phi_2$ and $\phi_-$ 
turn out to be
unmeasurably small, the method fails at step one.

Let us begin with the ratios of the four measured asymmetries: the 
total cross section differences and the enhanced parts of
$A_{NN}$ and $A_{LL}$. From these one can get immediately the
ratios of real to imaginary parts for $\phi_2$ and $\phi_-$ 
\begin{eqnarray} \label{eq:6.3}
\rho_2 &=& - 2 a_{NN}\,\sgmtot/P^2 \Delta \sigma_{\rm T}, \nonumber \\
\rho_- &=& 2 a_{LL}\,\sgmtot/P^2 \Delta \sigma_{\rm L}.
\end{eqnarray}
This fixes the phase of both amplitudes $\phi_2$ and $\phi_-$.
From the same four measurements a third independent ratio can be
formed; either 
\begin{equation}
\frac{P^2 \,\Delta \sigma_{\rm T}}{P^2 \, \Delta \sigma_{\rm L}} = -
\frac{I_2}{I_-}
\end{equation}
or
\begin{equation}
\frac{a_{LL}}{a_{NN}} = \frac{R_-}{R_2}
\end{equation}
will fix the ratios of the magnitudes of  $\phi_2$ and
$\phi_-$. We will use the latter in the following. 

In order to completely fix
the magnitudes, one more ratio is needed. Either
$b_{NN}/a_{NN}$ or
$b_{LL}/a_{LL}$ will do. Examination of the  table will reveal
that either of these quantities depends only on $I_2$ or,
equivalently, $I_-$ in addition to the ratios just determined;
the unknown $I_2$, say, is thereby related {\em linearly} to the
ratio
$b_{LL}/a_{LL}$ with known coefficients:
\begin{equation} \label{eq:I_2}
I_2 =  \frac{a_{LL}}{a_{NN}}
\frac{((b_{LL}/a_{LL}) - 1/\rho_- -\rho)}{\rho_2 +
1/\rho_2}.
\end{equation}

At this point, one has enough information to determine the polarization
because one can calculate $R_2, I_-$ and $R_-$ from Eq.\ (\ref{eq:I_2})
and the
previously determined quantities: one uses either $a_{LL}$ or 
$a_{NN}$ in
\begin{equation}
P^2 = \frac{a_{NN}}{R_2}
\end{equation}
or
\begin{equation}
P^2 = \frac{a_{LL}}{R_-}
\end{equation}
to obtain
\begin{equation}
P^2 = \frac{a_{NN}^2 + (P^2 \Delta \sigma_{\rm T}/2 \sgmtot)^2}{b_{LL} - P^2
(\Delta \sigma_{\rm L} 
/2 \sgmtot) - \rho a_{LL}}.
\end{equation}
This equation is valid in all the various degenerate limits except the case
$\phi_2 = 0$, both real and imaginary parts, in which case it is indeterminate
and one must work harder. 

Barring this exceptional case, one is in principle done because with this $P$
---presumably the sign ambiguity will not present a problem---one can use the
table to calculate $\phi_5$ from $A_N$; $A_{SL}$ is not needed. To give an
idea of the sensitivity of $A_{NN}$ to $R_2$, the curve for  $A_{NN}$ has
essentially the same shape as the CNI curve for $A_N$ and, for
$R_2 = 0.02$, the height at the maximum is about $2 \%$. It may very well
happen that $A_{NN}$ is measurable but that the error is too large for this to
provide a precision measurement; $\pm 1 \% $ would not be useful in the
example just cited. Here, too, one may benefit from pressing on: an error of 
$\pm 1 \% $ in $I_5$ would be far better than is required because it is
applied to a term of order 1 in $A_N$. 

Going further requires bringing in
$b_N/a_N$ and $b_{SL}/a_{SL}$. Each of these can be used to
express $I_5$ in terms of  $\rho_5$ and measured
quantities. By equating these two expressions an equation for
$\rho_5$  is obtained. The result is 
\begin{equation}\label{eq:rho_5}
\rho_5 = \frac{1}{b_N/a_N + b_{SL}/a_{SL}}
\left \{ \frac{(b_{SL}/a_{SL})((b_N/a_N) +
\rho + R_2)}{1 + I_2} -
\frac{(b_N/a_N)(I_2 +I_-)}{R_2 + R_-} \right\}.
\end{equation}
If this is then inserted into the equation for, say, $b_N/a_N$
then $I_5$ is determined since we have 
\begin{equation} \label{eq:I_5}
\frac{1}{I_5} = \frac{2}{\kappa (b_N/a_N +
b_{SL}/a_{SL})}
\left \{\frac{(b_N/a_N) + \rho + R_2}{1 +
I_2} +
\frac{I_2 +I_-}{R_2 + R_-} \right \}.
\end{equation}
Notice that there are no quadratic ambiguities in any of these determinations.
This is valid in all degenerate cases as well, as can be easly checked; it
only fails if both $\phi_2$ and $\phi_-$ vanish. The problem then becomes
identical to that of a proton scattering off a spin 0 particle for
which one cannot calculate the spin dependence without knowing $P$.

This procedure can be extended to apply to the case where the two spin $1/2$
particles are distinguishable, as in $p\ $-\ $^3{He}$ scattering. The part
concerning
$A_{NN},A_{LL}$ and $A_{SS}$ is identical. There are two new quantities to
determine, $\rho_6$ and $I_6$, and there are two additional equations,
effectively from $b_{SL}$ and $a_{SL}$ and from $b'_{N}$ and $a'_{N}$. These
can be solved just as in Eq.\ (\ref{eq:rho_5}) and Eq.\ (\ref{eq:I_5}). 

One can imagine a number of
special cases. An interesting case is pure pomeron pole dominance. In
that case (cf. Section 2)
$\phi_+, \, \phi_2$ and $\phi_5$ are all in phase while $\phi_- =
0$. In this very simple case, which should be easily checked
experimentally, 
$b_{SL}/b_{NN} = I_5$ and so $\phi_5$ is
determined in terms of measured quantities. Equivalently, one can
use the ratio of the hadronic piece to the enhanced piece of $A_{SL}$.
The corresponding ratio of the hadronic piece of $A_{LL}$ to the
enhanced piece of $A_{NN}$ determines
$I_2$ so everything is fixed:
\begin{eqnarray} I_5 & = & {\kappa \over
2}\frac{b_{SL}}{a_{SL}} \frac{1} {(\rho + 1/
\rho)},\nonumber \\ I_2 & = &  \frac{b_{LL}}{a_{NN}}
\frac{1}{(\rho + 1/\rho)}. 
\end{eqnarray}

One can easily take into account the Bethe phase corrections to this
procedure. Evidently, it will modify only the $a_i$ and has no effect on the
$b_i$. We have already seen in Section 3 that because $\delta$ is so small and
because it enters $A_N$ only by multiplying small quantities $\rho$ and $r_5$
or $r_2$, it can be safely neglected in $a_N$ to the accuracy that we are
working. The corrections to $A_{NN}$ and $A_{LL}$ are very similar; so, $a_{NN}
\to P^2 (R_2 + \delta \ I_2)$ and $a_{LL}
\to P^2 (R_- + \delta \ I_-)$ to lowest order in $\delta$; thus Eq.\
(\ref{eq:6.3}) becomes 
\begin{eqnarray} 
\rho_2 +\delta &=& - 2 a_{NN}\,\sgmtot/P^2 \Delta \sigma_{\rm T} \nonumber \\
\rho_- +\delta &=& 2 a_{LL}\,\sgmtot/P^2 \Delta \sigma_{\rm L}.
\end{eqnarray}
Since $\delta$ is a known quantity, the values of $\rho_2$ and $\rho_-$ can be
determined for use in the subsequent steps.

We now return to the $\alpha t$
corrections; these are small but they may need to be taken into account in
order to use this method if the amplitudes
$\phi_-, \phi_2$ and $\phi_5$ are quite small. The explicit
expressions for these terms are given in detail for all of these asymmetries
in \cite{BGL}. There are several sources of these corrections. The most
important arises from the slopes of the forward hadronic amplitudes, call them
$B_i$. In the purely hadronic part they appear only in order $t^2$ but,
through interference with the Coulomb singularities in either $\phi_+$ or
$\phi_5$, they contribute to $b_i$. It is very likely that the slopes for the
helicity-flip amplitudes are not very different from the non-flip $B_+$, a
factor of 2 at most; cf. the discussion in Section 4. To the degree that they
are the same the correction to $b_i$ is just
$a_i \,\alpha \, 4 \pi  B_+ /\sgmtot$. This corrects the corresponding
ratio $b_i/a_i$ by a known amount and can be simply accounted for. To the
degree that the slopes $B_i$ are different this procedure leaves behind a
correction of $(B_i - B_+)/2$ multiplied by one of the presumably small
amplitudes $\phi_2, \phi_-$ or $\phi_5$  and so is at a
level of about $10^{-3}$. The forward slopes of the Coulomb amplitudes can be
taken account of, in exactly the same way.

There is a correction to the real part of $\phi_2$ equal to $2 \alpha \kappa
^2/4 m^2$ which is about 0.01; this can simply be added into $R_2$ and
everything goes through as before. The term proportional to $\alpha t$ arising
from $|\phi_5|^2$ is of order $10^{-3}$ and so can be ignored. 

Finally there is the heretofore unmentioned $\phi_4$ which vanishes linearly
with $t$ as $t \rightarrow 0$.
Although the amplitude never enters the enhanced
piece, it does enter through interference into the linear term in $t$. One
guesses that its contribution will be negligible, but since nothing is
known about it, one would like to make sure that it can be controlled. Indeed,
it can in principle be determined by this method: this amplitude can be
removed from the first steps of the game by using 
$(A_{NN}+ A_{SS})/2$, instead of $A_{NN}$. The determination of the
amplitudes $\phi_-$ and $\phi_2$ goes through as before. Then by considering
$(A_{NN}-A_{SS})/2$ one can determine $R_4$. This can be used to correct
$A_{SL}$ which can in turn be used to fix $\rho_5$. Finally, then $b_N$ can be
used to fix $I_4$ and everything is determined. 

We don't want to oversell this method for self-calibrating CNI
polarimetry; we realize it is experimentally very uncertain. However, even if
the essential asymmetries are too small for this method to succeed, this linear
parametrization, making use of the CNI  enhancements, should prove useful for
determining the amplitudes at $t=0$, when the polarization is independently
measured. Furthermore, we find it interesting that it is possible, at least in
principle, to determine all of the spin dependent amplitudes without knowing
the beam polarization independently.

\section{Conclusions}
Motivated by the need to have an accurate knowledge of the proton-proton
single helicity-flip amplitude $\phi_5$ at high energies, in order to  devise
an absolute polarimeter for use in the forthcoming RHIC spin program, we
have examined the evidence for the existence of an asymptotic part of
$\phi_5$
which is not negligible compared to the largely imaginary average
non-flip amplitude
\(
         \phi_+  =  \case{1}{2}(\phi_1 + \phi_3)
\)
at high energies.  There is a general prejudice that
\(
        r_5 = m\,\phi_5/\s{-t}\,\im \phi_+
\)
will be negligibly small at high energies, say for
$
        p_L > 200
$
GeV/$c$, and we have tried, using various techniques,
to assess the validity of this belief.
We have explained how certain characteristics of the dynamical
mechanisms are linked to the behavior of the helicity amplitudes at
high energies and small momentum transfers, namely their growth with
energy, their phases, their small-$t$ behaviour, and relations amongst
them.
On the basis of rigorous analytical methods we have demonstrated that
the same fundamental assumptions which lead to the Froissart bound,
\(
        | \phi_+ | < s \ln^2 s \, ,
\)
permit $r_5$ to grow like $\ln s$. This surprising result implies that there is
nothing in principle to stop $\phi_5$ from remaining large, or even growing,
relative to $\phi_+$ at high energies.
However, other methods of analysis, based either on information at low
to medium energies, or based upon dynamical models, do suggest a small
$\phi_5$ at RHIC energies, typically $|r_5| < 15\%$.

Experimentally, for the region of interest to us, the best
constraint on $\phi_5$ comes from the measurement of $A_N$
in the CNI region at $ p_L = 200$ GeV/$c$.
Assuming that the phase of $\phi_5$ is the same as that of $\phi_+$
---
a sensible assumption for an asymptotically surviving contribution
---
one finds $ |r_5 | = 0.00 \pm 0.16$.  However, freeing the phase
yields
$       |r_5| = 0.2 \pm 0.3
$
and a phase difference between $\phi_5$ and $\phi_+$ of $0.15 \pm 0.27$ radians.
We
believe that the former value is the more reliable.
There are conflicting non-perturbative estimates of $r_5$  at $t = 0$.
By attempting to link helicity-flip to the isoscalar anomalous
magnetic moment of the nucleon, Landshoff and Polkinghorne arrive at
\(
        \im r_5 = -0.06 \, .
\)
This result is supported by an eikonal analysis of Bourrely,Soffer and collaborators,
who find
\(
        \im r_5 = -0.06
\)
when the nucleon matter density is taken equal to the charge density.
However, a more realistic choice of matter density leads to
\(
        \im r_5 = -0.018
\)
at $\s{s} = 50$ GeV and $-0.026$ at $\s{s} = 500$ GeV.
Surprisingly,
a study by the ITEP group, based upon the importance for
helicity-flip of the peripheral interaction with the pion cloud in
the nucleon, and which should therefore not be too different from
analyses based upon the matter density, yields
\(
        \im r_5 = 0.06
\)
i.e., of opposite sign to the above mentioned results.  On the other
hand Ryskin has attempted to calculate the anomalous colour magnetic
moment of a quark, based upon a mixture of perturbative QCD and the
constituent quark model, and linking the result to $\phi_5$ obtains
\(
        \im r_5 = 0.13
\)
i.e., of opposite sign to the results based upon the electromagnetic
anomalous moment.
Perturbative QCD attempts by Kopeliovich and Zakharov to link the existence of
helicity-flip to the transverse momentum of the consituents turn out to be very sensitive to
the form of the nucleon wave-function. If the wave function contains a significant
component corresponding to a compact scalar ($ud$) diquark they find that
$\im r_5$ increases in magnitude as the diquark size $D$ decreases.
Quantitatively
\(
        \im r_5 = -0.05 \to -0.15
\)
for $ D = 0.5 \to 0.2 \, $fm.

In summary while the various approaches give results which differ in
sign and magnitude, and while it is not clear to what extent
perturbative and non-perturbative approaches overlap, it seems
reasonable to assert that $ | r_5 | < 10\% $ at RHIC enrgies.
This level of accuracy is unfortunately inadequate for the needs
of an absolute polarimeter.
We have also studied the amplitudes
\(
         \phi_-  =  \case{1}{2}(\phi_1 - \phi_3)
\)
and $\phi_2$. There is persuasive evidence both from experiment
and from dynamical arguments that
$\phi_-$
is exceedingly small at high energies:
$
	|\phi_-/\phi_+| < 10^{-3}
$
for energies beyond $p_L = 200$ GeV/$c$. The case of $\phi_2$ is less
clearcut.
There is experimental evidence, but from relatively low energy
measurements of
$\delsigT$,
that
$ \im \phi_2$
drops from $-6\% \to -0.4\%$ for $p_L = 2 \to 6$ GeV/$c$.
And there is evidence from charge exchange scattering that the
$I=1$ part of
$\phi_2$
is very small at higher energies:
\(
	| r_2 | < 0.006
\)
at $p_L = 270$ GeV/$c$.
On dynamical grounds we expect $|r_2| \to 0$,
but the argument is not conclusive.

Finally, we have demonstrated the surprising result that
proton-proton elastic scattering is self analysing,
in the sense that all the helicity amplitudes can be
determined experimentally at very small momentum transfer, without a
knowledge of the magnitude of the beam and target polarization.
The experimental procedure for doing this is complex, but once carried
out successfully it would permit the calibration of a CNI polarimeter
which could then be used very simply for routine measurement of the beam
polarization.

{\bf Acknowledgements} We would like to thank the following people for
significant discussions: N. Akchurin, E. Berger, C. Bourrely, G. Bunce, W.
Guryn, A. Krisch, P.\ Landshoff, S. MacDowell, A. Penzo, T. Roser and O.V.
Selyugin. We would like to give special thanks to Y. Makdisi for raising
questions that led to this work and to the RIKEN-BNL Research Center for
sponsoring a workshop at which this work was begun.


\begin{thebibliography}{99}
\bibitem{HERA} H1 Collab., S.\ Aid et al., Nucl.Phys. {\bf B470}, 3 (1996),
ZEUS Collab., M.\ Derrick et al., Z.~Phys. {\bf C72}, 399 (1996),
A.~H.\ Mueller, Eur.Phys.J. {\bf A1}, 19 (1998).
\bibitem{RHIC} Y.~I.\ Makdisi, {\em Polarization in Hadron-Induced Processes at
RHIC}, Proceedings of ``Spin 96'', 12th
Int. Symp. on High Energy Spin
Physics, Amsterdam Sept.96, Eds. C.W. de
Jager, et al., World
Scientific, Singapore, 1997, p.107. 

\bibitem{Guryn}	W. Guryn {\it et al}.,  {\it PP2PP\/} Proposal to Measure
	Total and Elastic $pp$ Cross Sections at RHIC (unpublished).
\bibitem{schwinger}
        B.~Z. Kopeliovich and L.~I.
Lapidus, Sov.~J.~Nucl.~Phys.{\bf 19}, 114 (1974).
\bibitem{BGL}
N.~H. Buttimore, E. Gotsman, and E. Leader,
Phys.~Rev.{\bf D 18}, 694 (1978).
\bibitem{kz} 	B.~Z.~Kopeliovich and
B.~Z.~Zakharov, Phys.~Lett. {\bf B 226}, 156
(1989).
\bibitem{trueman}
	T.~L.~Trueman, {\em CNI Polarimetry and the hadronic spin
	dependence of p p scattering,} 
  Proceedings of ``Spin 96'', {\it op. cit.}, pp. 833,
 (hep-ph/9610429); 
	RHIC/DET Note 18 (1996), (hep-ph/9610316).
\bibitem{bs} C. Bourrely and J. Soffer, {\em How to calibrate the polarization
of a high energy proton beam? A theoretical prospect}, Spin 96, {\it op. cit.}
pp. 825.
\bibitem{polarimetry requirement} C. Prescott et al, Report of RHIC Spin Review
Committee, June 1995 : ``The Committee feels the physics goals of STAR and
PHENIX require a polarimeter system accurate to $\Delta P/P = \pm 5 \%$
(absolute calibration, not just relative calibration).''
\bibitem{Amaldi} U. Amaldi et al,
Phys.~Lett. {\bf 43B}, 231, (1973).

\bibitem{BSW75} C.~Bourrely, J.~Soffer and D.~Wray, Nucl.~Phys. 
{\bf B91}, 33 (1975)
\bibitem{bsw}
C. Bourrely, J. Soffer, and D.~Wray,  Nucl.~Phys. {\bf B77}, 386 (1974).
\bibitem{Martin/Marseille} A.
Martin, Jour.~de~Phys. {\bf 46}, C2-727 (1985). 

\bibitem{RIKEN} {\em Hadron Spin-flip at RHIC Energies}, Proceedings of RIKEN
BNL Research Center Workshop, Vol.3, 1997 (BNL 64672). 
\bibitem{jet} N. Akchurin et al, {\em An Absolute 
and Non-destructive Polarimeter for HERA-p}, Spin 96 {\em op. cit.}, p. 804. 
\bibitem{ep} I.~V. Glavanakov et al, {\em Elastic $pe$-scattering 
as analyzer of high energy proton beams polarization}, Spin 96 {\em op. cit.},
p.794.
\bibitem{BZK1} B.~Z. Kopeliovich, {\em High Energy Polarimetry at RHIC}, MPI H-V3-1998
(hep-ph/9801414).
\bibitem{bs2} C. Bourrely and J. Soffer, Phys.~Lett.~{\bf B442}, 479 (1998).
\bibitem{KT} B.~Z. Kopeliovich and T.~L. Trueman, to be published.
\bibitem{inclusive} I. Alekseev et al, {\em Conceptual Design of a Proton
Polarimeter for RHIC}, \\ Spin 96 {\em op. cit.}, p.797.

\bibitem{Gribov} V.N. Gribov, Soviet~Journal~of~
Nuclear Physics {\bf 5}, 138 (1967).
\bibitem{MT1} A.~H. Mueller and T.~L.
Trueman, Phys.~Rev. {\bf 160}, 1296 (1967).
\bibitem{10} E. Leader and R. Slansky,
Phys.~Rev. {\bf 148} (1966) 1491.
\bibitem{Volkov} D.~V. Volkov and V.~N.
Gribov, JETP {\bf 44}, 1068 (1963), Soviet
Physics, JETP {\bf 17}, 720 (1963).

\bibitem{GGMW}
M.~L. Goldberger, M.~T. Grisaru, S.~W. MacDowell and D.~Y. Wong,
Phys.~Rev. {\bf 120}, 2250 (1960).
\bibitem{cahn}
R.~N. Cahn, 	Z.~Phys., {\bf C15}, 253 (1982) .
\bibitem{predazzi}
	E.~Leader and E.~Predazzi,
{\it Gauge Theories and the `New
Physics'}, Cambridge University Press (1982).
\bibitem{Gauron} P. Gauron, B.
Nicolescu and E. Leader, Nucl.~
Phys. {\bf B299}, 640 (1988).
\bibitem{Block} M.~M. Block and R.~N. Cahn, Phys.~Lett. {\bf 168B}, 151
(1986), M.~M. Block, B. Margolis, and A.~R. White, hep-ph/9510290.
\bibitem{1}  M. Froissart, Phys.~Rev.
{\bf 123}, 1053 (1961); A. Martin, Nuovo~Cimento {\bf 42}, 930 (1966); {\em
ibid.}{\bf 44}, 1219 (1966).

 
\bibitem{2} A. Donnachie and P.~V.
Landshoff, Nucl.~Phys. {\bf B244}, 322 (1984); {\em ibid} {\bf B267}, 690
(1986); Phys.~Lett. {\bf B185}, 403 (1987).
\bibitem{3} L.~N. Lipatov, Pomeron in
QCD, in: {\sl Perturbative QCD},
A.H.~Mueller (ed), World Scientific,
Singapore (1989). 
\bibitem{chengwu} H. Cheng, J.~K. Walker, T.~T. Wu, Phys.~Lett.
{\bf B44}, 97 (1973).
\bibitem{Volkovitsky} P.~E.~Volkovitsky, A.~M.~Lapidus, V.~I.~Lisin and
K.~A.~Ter-Martirosyan, Sov.~J.~Nucl.~Phys. {\bf 24}, 648 (1976).
\bibitem{Dubovikov} M.~S. Dubovikov,
B.~Z. Kopeliovich, L.~I. Lapidus, and
K.~A. Ter-Martirosyan, Nucl.~Phys.
{\bf B123}, 147 (1977). 
\bibitem{5} See e.g. P. Gauron,
L.~N.~Lipatov and B. Nicolescu, Phys.~Lett. {\bf B304}, 334 (1993). For more
recent calculations indicating that the odderon lies slightly below $J=1$ at
$t=0$, see J. Wosiek and R.~A. Janik, Phys.~Rev.~Lett. {\bf 79}, 2935 (1997)
and N. Armesto and M.~A. Braun, Z.~ Phys. {\bf C75}, 709 (1997).
\bibitem{6} P. Gauron, E. Leader and
B. Nicolescu, Phys.~Lett. {\bf B238}, 406 (1990). The odderon
was first proposed in L. \L ukaszuk and B.~Nicolescu, Il Nuovo
Cim. Lett.{\bf 8}, 405 (1973)  and named in D. Joynson, E. Leader and B. Nicolescu,
Il Nuovo Cimento {\bf 30}, 345 (1975).

\bibitem{11} E. Leader, Phys.~Rev. {\bf 166}, 1599 (1968) .
\bibitem{PT} R.~F. Peierls and T.~L. Trueman, Phys.~Rev. {\bf 134}, B1365
(1964).
\bibitem{Eden} R.~J. Eden, Rev.~Mod.~Phys. {\bf 43}, 15 (1971).
\bibitem{s-channel} F.~J. Gilman, J. Pumplin, A. Schwimmer and L. Stodolsky,
Phys.~Lett.{\bf 31B}, 387 (1970), H. Harari and Y. Zarmi, Phys.~Lett. {\bf
32B}, 291 (1970).
\bibitem{7} E.~Leader and T.~L.~Trueman, to be published.
\bibitem{Mandelstam} S. Mandelstam, Il Nuovo Cimento {\bf 30}, 1113, 1117, 1148
(1963).

\bibitem{Gribov1} V.~N. Gribov,
JETP (Sov.Phys.) {\bf 26}, 414 (1968).
\bibitem{Polkinghorne} J.~C.
Polkinghorne, Nuovo~Cimento {\bf 56A}, 755
(1968).
\bibitem{Branson} D. Branson,
Phys.~Rev. {\bf 163}, 1608 (1969).
\bibitem{Jones} L.~M. Jones and P.~V.~Landshoff,
 Nucl.~Phys. {\bf B94}, 145 (1975).

\bibitem{E704}  N. Akchurin et al.,
Phys.~Lett. {\bf B229}, (1989);  Phys.~Rev. {\bf D48}, 3026 (1993).
\bibitem{nhb-bnl} N.~H. Buttimore, AIP Conf. Proc. No.~95,
High Energy Spin Physics, Brookhaven, 1982, ed.~G.~M. Bunce 
(AIP, New York, 1983), p.~634.

\bibitem{abp} N. Akchurin, N.~H.
Buttimore and A. Penzo, 
Phys.~Rev. {\bf D51}, 3944 (1995);
Proceedings of the VIth Blois Workshop, Blois,
20-24 June 1995, Edition Fronti\`eres 1996, p.~411.

\bibitem{WdeB} W. de Boer et al.,
Phys.~Rev.~Lett. {\bf 41}, 558 (1975);
E. K. Biegert et al.,
Phys.~Lett. {\bf 73B}, 235 (1978).
\bibitem{FI} G.~P. Farmelo and A.~C.
Irving, Nucl.~Phys. {\bf B128}, 349
(1977). 
\bibitem{auer} I.~P. Auer et al., Phys.~Lett. {\bf 70B}, 475 (1977),
Phys.~Rev.~Lett. {\bf 62}, 2649 (1989), Phys.~Rev. {\bf D55}, 1159 (1997).
\bibitem{Grosnick} D.~P. Grosnick et
al., Phys.~Rev. {\bf D55}, 1159 (1997) .
\bibitem{krisch}
 A.~D.~Krisch and S.~M.~Troshin, {\em
 Estimate of elastic proton-proton
polarization at small $P_T^2$ near 1
TeV,}
 Proceedings of ``Spin 96'',{\em op. cit.}, p.830.
\bibitem{data}  
M.~Borghini et al., Phys.~Lett. {\bf 36B}, 501 (1971),
S.~L.~Kramer et al., Phys.~Rev. {\bf D17}, 1707 (1977),
D.~G.~Crabb et al., Nuc.~Phys. {\bf B121}, 231(1977),
A.~Gaidot et al., Phys.~Lett. {\bf 61B}, 103(1976),
J.~H.~Snyder et al., Phys.~Rev.~Lett. {\bf 41}, 781(1978).
\bibitem{i&w} A. Irving and R. Worden, Phys.~Rep. {\bf 34C}, 117 (1977).
\bibitem{PDB} P.~D.~B.~Collins, {\it An
introduction to Regge theory \& high
energy physics}, Cambridge University
Press, Cambridge, 1978; P.~D.~B. Collins and P.~J. Kearney, Z.~Phys. {\bf
C22}, 277 (1984).
\bibitem{berger} E.~L. Berger, A.~C. Irving,
and C. Sorensen, Phys.~Rev. {\bf D17}, 2971
(1978).
\bibitem{bls}
C. Bourrely, E. Leader, and J. Soffer,
Phys.~Reports {\bf 59}, 95 (1980).
\bibitem{ira1} I.~K.~Potashnikova, Sov.~J.~Nucl.~Phys.
{\bf 2}, 674 (1977). 

\bibitem{tot} R.~M.~Barnett et al.,
Phys.~Rev. {\bf D54}, 1 (1996). 

\bibitem{b67} M.~Borghini et al.,
Phys. Lett. {\bf B24}, 77 (1967). 
\bibitem{b70} M.~Borghini et al.,
Phys. Lett. {\bf B31}, 405 (1970). 
\bibitem{b71}M.~Borghini et al. Phys. Lett. {\bf B36}, 493 (1971).

\bibitem{JW} M. Jacob and G.~C. Wick, {\it
Ann.~Phys.} {\bf 7}, 404 (1959).
\bibitem{AG} M. Andrews and J. Gunson, J.~Math.~Phys. {\bf 5},1391 (1964).
\bibitem{MS} A.~D. Martin and T.~D. Spearman, {\it Elementary Particle
Theory}, North-Holland Publishing Co., Amsterdam, 1970.   
\bibitem{mutter} K.~H. M\"{u}tter, 
Nucl.~Phys. {\bf B27}, 73 (1971), {\bf B31}, 589 (1971).
\bibitem{hodgkinson} D.~P.~Hodgkinson,
Phys.~Lett.~{\bf 39B}, 640 (1972).
\bibitem{mennessier}
 G.~Mennessier, S.~M.~Roy, and V.~Singh,
 Nuovo Cimento, {\bf 50A}, 443 (1979);
 K.~S.~Ramadurai and I.~A.~Sakmar,
 Prog.~Th.~Phys.~{\bf 63}, 1700 (1980).
\bibitem{as} U.~Amaldi, K.~R.~Schubert,
Nucl.~Phys. {\bf B166}, 301 (1980). 


\bibitem{lands&polk} P.~V. Landshoff and J.~C. Polkinghorne, Nucl.~Phys. {\bf
B32}, 541 (1971).
\bibitem{ryskin}
M.~G. Ryskin, Yad.~Fiz. {\bf 46}, 611 (1987); 
Sov.~J. Nucl.~Phys. {\bf 46}, 337 (1987).
\bibitem{Shuryak} E.~V.~Shuryak, Rev.~Mod.~Phys. {\bf 65}, 1 (1993).
\bibitem{triple-pom} M.~Genovese et al., J.~Exp.~Theor.~Phys. {\bf 81}, 633
(1995).
\bibitem{z} B.~G.~Zakharov, Sov. J.~Nucl.~Phys. {\bf 49}, 860 (1989). 
\bibitem{kp}  B.~Z.~Kopeliovich and
B.~Povh,Modern Physics Letters {\bf A13}, 3033 (1998).
\bibitem{knp} B.~Z.~Kopeliovich, N.~N.~Nikolaev and I.~K.~Potashnikova,
Phys.~Rev. {\bf D39}, 769 (1989).
 \bibitem{pumplin} J.~Pumplin and
G.~L.~Kane, Phys.~Rev. {\bf D11}, 1183 (1975) 
\bibitem{pion} K.~G.~Boreskov,
A.~A.~Grigiryan, A.~B.~Kaidalov  and
I.~I.~Levintov, Sov.~J.~Nucl.~Phys.
{\bf 27}, 432 (1978). 
\bibitem{regge} A.~M. Lapidus and P.~E.
Volkovitskii, Sov.~J.~Nucl.~Phys. 
{\bf 31}, 380 (1980).

\bibitem{BSW79} C.~Bourrely, J.~Soffer and T.~T.~Wu, 
Phys.~Rev. {\bf D19}, 3249 (1979).
\bibitem{BSW84} C.~Bourrely, J.~Soffer and T.~T.~Wu, Nucl.~Phys. 
{\bf B247}, 15 (1984).
\bibitem{BSW95} C.~Bourrely, J.~Soffer and T.~T.~Wu, Proceedings of the VIth
Blois Workshop, Blois, 20-24 June 1995, Editions Fronti\`eres 1996, pp.15 and 
references therein. 
\bibitem{B82} C.~Bourrely, H.~A.~Neal, G.~A.~Ogren, J.~Soffer and T.~T.~Wu,
Phys.~Rev. {\bf D26}, 1781 (1982). 


\end{thebibliography}
\end{document}